# In-Circuit Characterization of Low-Frequency Stability Margins in Power Amplifiers

Jose Manuel Gonzalez, Nerea Otegi, Aitziber Anakabe, Libe Mori, Asier Barcenilla, and Juan-Mari Collantes

***Abstract*— Low-frequency resonances with low stability margins affect video bandwidth characteristics of power amplifiers. In this work, a non-connectorized measurement technique is presented to obtain the low-frequency critical poles at internal nodes of a hybrid amplifier. The experimental setup uses a high impedance probe connected to a vector network analyzer (VNA) to obtain a fully calibrated closed-loop frequency response that is identified to get the poles of the device at low frequency. Compared to previous connectorized solutions, the approach avoids the ad-hoc insertion of extra RF connectors to access the low-frequency dynamics of the amplifier. In addition, it simplifies the characterization at multiple internal nodes, which is worthwhile for an efficient detection and fixing of critical low frequency dynamics in multistage power amplifiers. The technique is first applied to dc steady state regimes and compared to the connectorized approach on a single stage amplifier. Next, it is applied to a three-stage amplifier to show its potential to detect the origin of the undesired dynamics and the most effective way to increase stability margin. Finally, the technique has been extended to the large-signal case to increase its usefulness for the design and diagnosis of high power amplifiers.**

***Index Terms*— Amplifier testing, high impedance probing, in-circuit characterization, poles and zeros, stability analysis.**

## I. INTRODUCTION

MOBILE communication systems have a continuous need to increase bandwidth in order to accommodate more channels at higher data rates. Power amplifiers are critical elements in the quest for large bandwidth/low distortion communication systems. The internal low-frequency dynamics of the amplifier has a direct impact on its video bandwidth as well as on the ability of the digital pre-distortion systems to correct for intermodulation distortion (to meet regulation standards) in signals with large instantaneous bandwidths [1], [2].

In this context, a limiting factor for the amplifier can be the presence of a pair of low-frequency complex conjugate critical poles that shift with some circuit parameter. These poles are often generated by the parasitic Hartley topology created between gate and drain bias paths and the intrinsic gate to drain capacitance [3]. When a parameter of the circuit is varied, such as temperature, bias or loading conditions, these poles may undergo a considerable shift on the complex plane. In the worst case scenario they can cross to the right-half plane (RHP) and become unstable, generating an undesired low-frequency oscillation. Even if the pair of poles does not become unstable, a position too close to the RHP implies low damping and a high resonant effect. This small stability margin has several negative effects: it increases the risk of oscillation making the amplifier less robust to parameter changes [4], [5], it is responsible for the presence of noise bumps in the output spectra [6], it affects transient response [7], and, as mentioned above, it limits the performance of the digital pre-distortion systems [1], [2].

In principle, the stability margins associated with these low-frequency critical poles may be evaluated in simulation. However, to get consistent results versus bias, temperature or input power, we need reliable models of the active devices, which sometimes may not be accurate enough [8], [9] or are not even available. In those circumstances an experimental method to characterize the stability margin is helpful either to complement/confirm simulation analyses or to substitute them when simulation is not a reliable option.

A technique to detect critical low-frequency resonances using coaxial probes was proposed in [10]. In [8], an experimental methodology was presented to measure and control low-frequency stability margins. The technique required the inclusion of extra RF ports at gate and drain bias paths to access the low-frequency dynamics. Then, from connectorized measurements of reflection coefficients at those

Manuscript submitted June 4, 2018; revised September 19, 2018. This work was supported by the Spanish Ministry of Economy and Competitiveness and the European Regional Development Fund (MINECO/FEDER) through Research Project TEC2015-67217-R, and by Basque Country Government through Project IT1104-16.

J.M. Gonzalez is with Departamento de Tecnología Electrónica, University of the Basque Country (UPV/EHU), Bilbao, 48013, Spain (e-mail: josemanuel.gonzalezp@ehu.eus)

N. Otegi. A. Anakabe, L. Mori, A. Barcenilla and J.M. Collantes are with Departamento de Electricidad y Electrónica, University of the Basque Country (UPV/EHU), Leioa, 48940, Spain (e-mail: nerea.otegi@ehu.eus; aitziber.anakabe@ehu.eus; libe.mori@ehu.eus; asier.barcenilla@ehu.eus; juanmari.collantes@ehu.eus).

ports, the position and evolution of low-frequency critical poles on the complex plane were obtained. However, this technique has a basic limitation because the presence and arrangement of the extra ports has to be planned and included in the layout of the prototype from the beginning of the design. If the low frequency characterization has not been anticipated at the design stage, the technique is inapplicable. Besides, multistage power amplifiers would require the inclusion of observation RF ports at gate and drain bias paths of each stage, making the layout of the prototype and the technique itself excessively cumbersome.

In this work we propose a non-connectorized solution to obtain the position of the low-frequency critical poles at internal nodes of a hybrid amplifier without the need of extra RF connectors. The technique is based on using a high impedance probe connected to a vector network analyzer (VNA) to access the internal nodes of the prototype. A procedure is given to obtain a fully calibrated closed-loop frequency response (in magnitude and phase) that represents the linearization of the circuit around its steady state regime. From this frequency response, the critical poles are identified using conventional frequency domain identification techniques for linear systems. Respect to the connectorized technique in [8], the approach here presented has the benefit of enabling the measurements at different internal stages of a multistage amplifier. The information extracted from these measurements can be used to determine where in the amplifier the origin of the critical dynamics is located and, consequently, where is more effective to act in order to correct this behavior.

Other non-connectorized approaches have been recently presented in the literature for diagnosis and characterization of microwave amplifiers and devices [9], [11]-[13]. Most of them have different goals than the one presented in this work. In [9], authors propose a technique to experimentally obtain bifurcation curves, including unstable states, imposing a null condition at a node of the circuit. The approach adapts the Auxiliary Generator technique [4] to the experimental environment. The probe that performs the null condition imposes a large-signal waveform corresponding to the actual steady-state at the injection node (for non-perturbation). In [11]-[13] near field electrical probes are used to obtain the time-domain waveforms of the large-signal steady state in various amplifiers. The aim of these works being a large-signal steady-state waveform, no linearization is performed. Our approach, by contrast, uses a small-signal source that is swept in frequency to linearize the steady-state regime (either dc or periodic) because we are seeking to obtain the poles of the system.

Different techniques for contactless characterization of scattering parameters are presented in [14]-[16]. In [14], [15] electromagnetic simulations of known passive structures embedded in the circuit are required to complement the measurements. In [16] scattering parameters are measured with contactless loop probes located along different positions of a microstrip transmission line. Although valuable for internal characterization, both types of techniques are difficult to apply to generic internal nodes of an amplifier. On the contrary, the proposed methodology is specifically suited for characterization at internal nodes of amplifiers whose layout has not been previously designed or arranged for that characterization.

The paper is organized as follows. Section II describes the in-circuit characterization technique required to obtain a closed-loop frequency response valid for identification at an internal node. Experimental validation of the technique is provided in Section III where results are compared to those obtained through the connectorized method in a single stage amplifier with extra RF connectors at the bias paths. The in-circuit characterization of a multistage amplifier is given in Section IV, showing the ability of the technique to measure the pole evolution and its capability to detect the critical zones of the amplifier responsible for the undesired dynamics. In Section V, the methodology is extended to large-signal steady states driven by the input power. Eventually, limitations of the technique and final discussions are given in Section VI.

## II. Methodology

In this section, the methodology to obtain the critical low-frequency poles at internal nodes of the circuit is explained.

In general, a practical way to obtain the poles of a linear (or linearized) system consists in applying pole-zero identification algorithms to a closed-loop frequency response of the system. In simulation, a common approach to obtain the closed-loop frequency response is to inject a small-signal current source at a particular node of the circuit, which will act as input variable, and to use the voltage at that node as output variable. Thus, the closed-loop frequency response is given by the impedance seen by the small-signal current source [17], [18]. Alternatively, and depending on the circuit configuration, it is also frequent to obtain the closed-loop frequency response by injecting a small-signal voltage source in series at a branch of the circuit, acting as input variable, and to monitor the current flowing through that branch as output variable. In this case, the closed-loop frequency response is given by the admittance seen by the small-signal voltage source.

Neither of these two approaches, commonly used in simulation, is well-suited for the experimental in-circuit characterization of the system poles due to the difficulty of injecting and measuring at the same internal point. Therefore, a transmission closed-loop frequency response is proposed instead. The technique is based on injecting a small-signal voltage source $v_{gen}$ in series at the input port of the circuit as input variable and obtaining the voltage $v_n$ at some internal node $n$ as output variable (Fig. 1). By doing so, the closed-loop frequency response $H_{vn}$ given by (1) represents now a voltage transmission. Note that, the system being linearized, all closed-loop transfer functions share the same poles.

$$H_{vn} = \frac{v_n}{v_{gen}} \tag{1}$$

In the experimental setup the technique is implemented by means of a VNA and a high impedance probe. The small-signal voltage source $v_{gen}$ is synthesized by the signal generator of the VNA whose port 1 is connected to the input port of the device under test (DUT) (Fig. 2). The high impedance active probe, connected to port 2 of the VNA as shown in Fig. 2, allows sensing at an internal node $n$ to obtain the voltage transmission response. It is indispensable to ensure that the probe is not affecting the circuit behavior during characterization. The high input impedance of the active probe must therefore guarantee this non-perturbation condition at least for the low-frequencies involved in the characterization. A discussion on the limits of this non-perturbation condition is developed in Section VI.

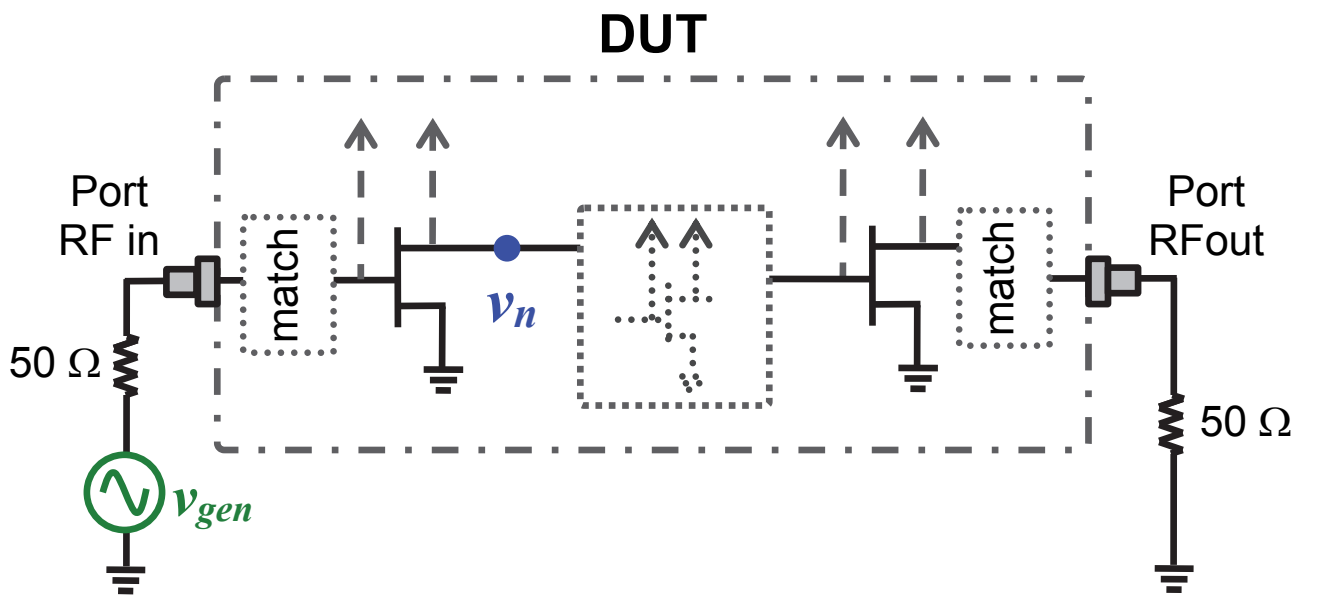

Fig. 1. Basic schematic of a circuit with small-signal voltage generator $v_{gen}$ as input variable and voltage $v_n$ at internal node $n$ as output variable. The voltage transmission closed-loop frequency response is $H_{vn}=v_n/v_{gen}$.

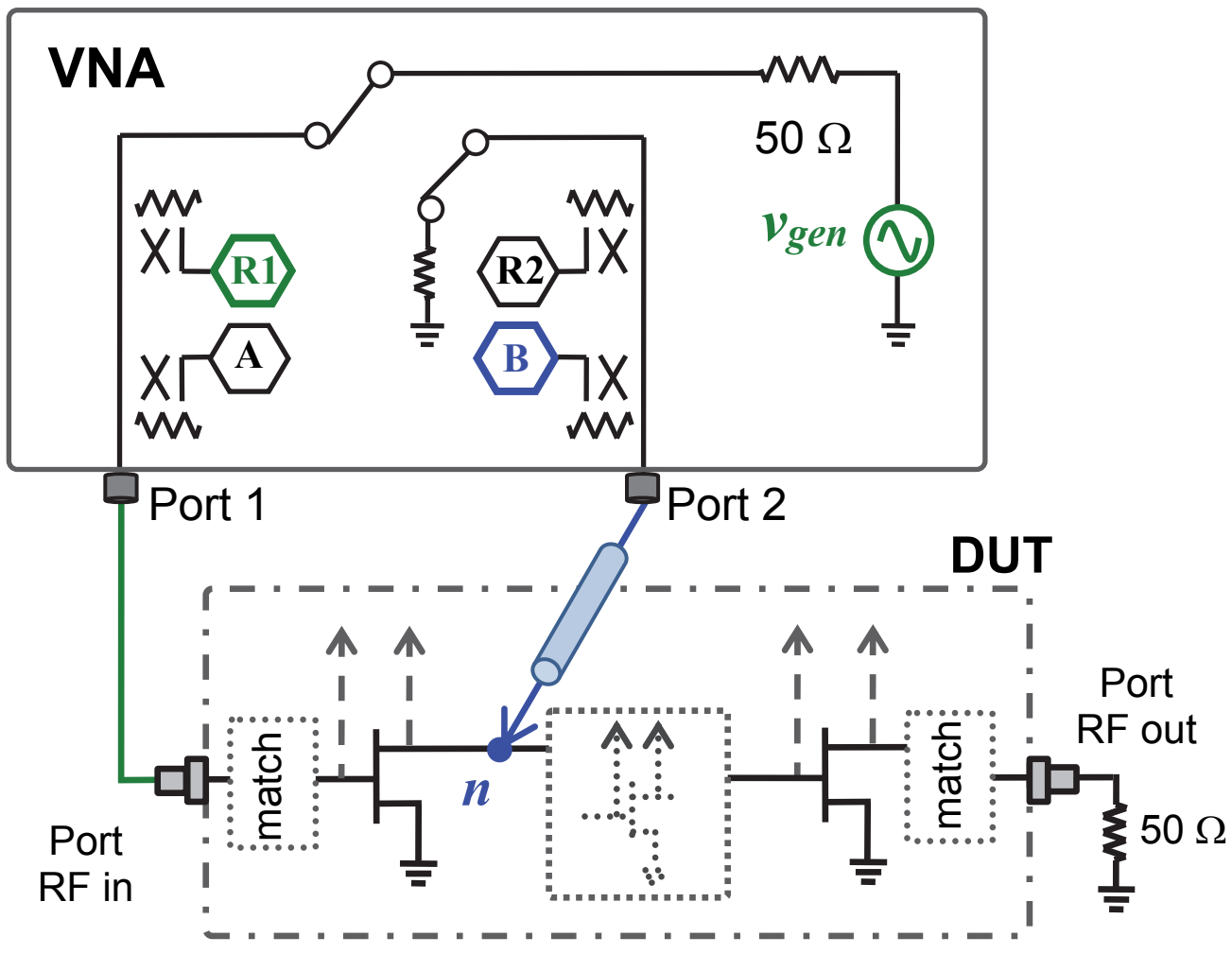

Fig. 2. Schematic of the measurement setup for the experimental implementation: port 1 of VNA is connected to DUT input and an active high impedance probe, connected to port 2 of VNA, senses at an internal node $n$.

In practice, computing $H_{vn}$ in the experimental setup is not direct. Using the high impedance probe, only voltage transmission measurements between two internal nodes of the circuit can be obtained [19]. However, we need to relate the voltage at the internal node $n$ with the small-signal voltage source $v_{gen}$ applied from port 1 of the VNA, which is obviously outside the DUT (Fig. 2). To do so, we consider $H_{vn}$ as the cascade connection of two voltage transfer functions, $H_n$ and $H_{input}$, that are calculated separately:

$$H_{vn} = H_{input} H_n \tag{2}$$

$H_n$ (3) is the voltage transfer function between node $n$ and an internal reference node. This reference node, called "input reference node", is located on the input transmission line, as close as possible to the input connector, as shown in Fig. 3.

$$H_n = \frac{v_n}{v_{ref}} \tag{3}$$

$H_{input}$ (4) is the voltage transfer function between the input reference node and the small-signal generator voltage $v_{gen}$ (Fig. 3).

$$H_{input} = \frac{v_{ref}}{v_{gen}} \tag{4}$$

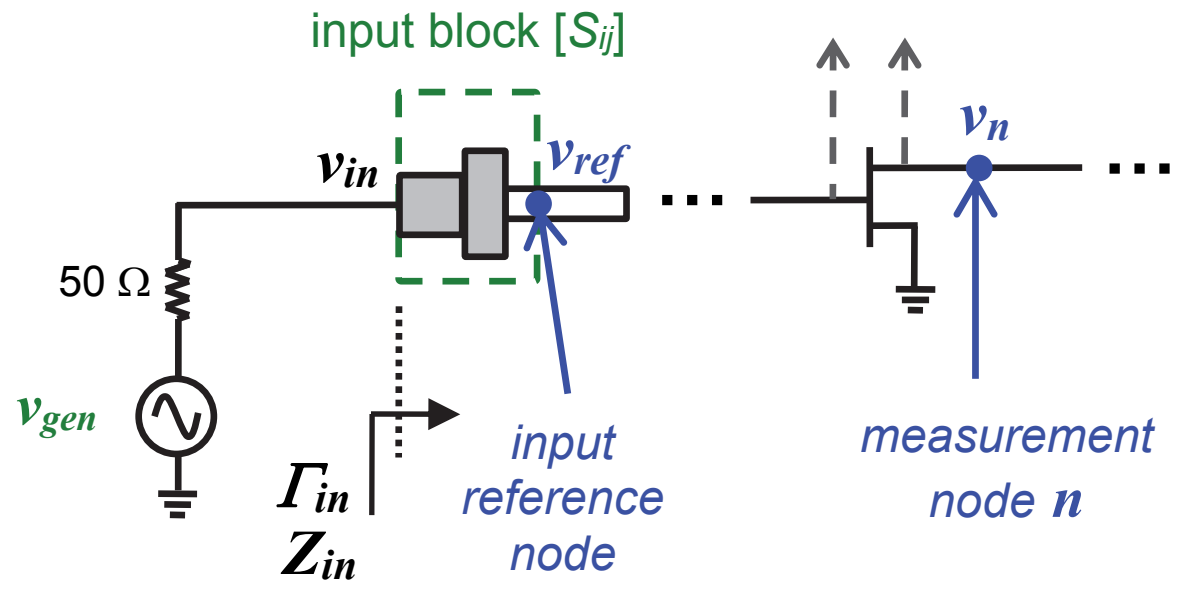

Fig. 3. Basic schematic showing input block from circuit input port to internal input reference node, immediately after input connector.

In the next two sub-sections the procedures to compute the two voltage transfer functions $H_n$ and $H_{input}$ are detailed.

## A. Calculation of $H_{input}$

To compute voltage transfer function $H_{input}$, a de-embedding process is necessary using conventional S-parameter measurements. This de-embedding accounts for the effect of the input block made up by the input connector in series with the short section of transmission line that goes from the connector to the position of input reference node (Fig. 3). In most cases, a simple ideal delay could be enough to model the input block because the characterization frequencies are typically low. However, for a more accurate de-embedding at higher frequencies, the scattering parameters of the input block have to be taken into account.

The voltage transfer function $H_{input}$ can be calculated from:

$$H_{input} = \frac{v_{ref}}{v_{in}} \frac{v_{in}}{v_{gen}} \tag{5}$$

where, considering that port 1 impedance of the VNA is 50 Ω:

$$\frac{v_{in}}{v_{gen}} = \frac{Z_{in}}{Z_{in} + 50\Omega} \tag{6}$$

and:

$$\frac{v_{ref}}{v_{in}} = \left( S_{21} + \frac{(\Gamma_{in} - S_{11})(1 + S_{22})}{S_{12}} \right) \frac{1}{1 + \Gamma_{in}} \tag{7}$$

$\Gamma_{in}$ and $Z_{in}$ are, respectively, the standard input reflection coefficient and input impedance of the DUT; and $S_{ij}$ are the S-parameters of the input block.

The input block can be characterized separately by fabricating a specifically designed second test device. This test device is a simple two-port device with two identical connectors linked by a short section of microstrip line (Fig. 4).

The device must have the same substrate, same kind of connectors and same transmission line as the input block of the amplifier under study. The middle plane of this device coincides geometrically with the position of the input reference node in the amplifier (Fig. 3). Therefore, apart from technological dispersion, the amplifier input block is similar to half of the test device. Through a two-port S-parameter characterization of the two-port test device, and using ABCD-matrix, the scattering parameters of one of the two halves can be approximated, assuming that the two halves are identical. In this way we have an accurate estimation of the scattering parameters $S_{ij}$ of the input block of the actual amplifier.

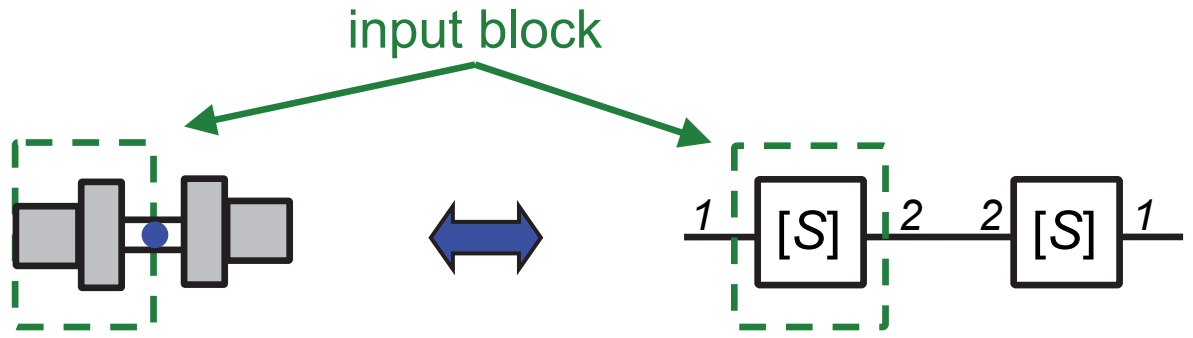

Fig. 4. Diagram of test device specifically fabricated to obtain the S-parameters of the input block [$S_{ij}$]. The device can be seen as the cascade of two symmetrical input blocks, so that the input block can be characterized by the bisection of the total device.

### B. Calculation of $H_n$

In order to obtain $H_n$, two measurements with the high-impedance probe are required. First, a thru calibration with the probe at the input reference node is performed, as explained in [19]. Next, probing at node $n$ directly provides the voltage transfer function $H_n$ between $v_n$ and $v_{ref}$. Actually, this calibrated measurement is analogous to performing two transmission measurements $B/R1$ (Fig. 2), one at the measurement node $n$ and the other at the input reference node, and then computing the ratio of them as in (8).

$$H_n = \frac{\left(B_n / R1_n\right)}{\left(B_{ref} / R1_{ref}\right)} \tag{8}$$

### C. Selection of the probing nodes

Impedances seen at sensing nodes should be significantly lower than the impedance of the probe to guarantee the non-perturbation condition. This is not a challenge for the input reference node because its impedance will be typically 50 Ω or lower. The selection of node $n$ has to be more careful. In principle, from a sensitivity point of view, placing the probe at the gate or drain nodes of the transistors may seem a reasonable option because it guarantees that we are sensing inside the loop responsible for the critical resonance. However, in practice, gate and drain impedances can be significantly high at video frequencies, which increases the loading of the high-impedance probe. Therefore, for low-frequency critical resonance detection, it may be more judicious to use nodes located on the bias paths, immediately after the bias inductance or quarter lambda transformer. It is also preferable to use nodes on the gate bias paths than on the drain bias paths because current and voltage levels are lower at the gate, which reduces the risk of damaging the probe. In general, bias line nodes at which RF signal has been decoupled are interesting because they can also be used for the large-signal characterization, as will be discussed in Section V. Further comments on the non-perturbing condition of the probe are given in section VI.

### D. Summary of the technique

To summarize, the basic steps to obtain the critical low-frequency poles are the following:

*1)* Select an internal node $n$ for observation.
*2)* Obtain the S-parameters of the input block as explained in Section II.B.
*3)* Characterize the input reflection coefficient $\Gamma_{in}$ of the amplifier. Then calculate $H_{input}$ from (5).
*4)* Perform a first measurement with the probe placed at input reference node.
*5)* Perform a second measurement with the probe at node $n$ to calculate $H_n$ from (8).
*6)* Compute $H_{vn}$ using (2).

After the closed-loop frequency response $H_{vn}$ has been determined, pole-zero identification techniques [17], [18] can be applied to obtain the associated pole-zero map.

## III. Experimental Setup and Validation

### A. Experimental setup

Our measurement setup for the in-circuit characterization of critical poles makes use of a high impedance active probe (Keysight 85024A) connected to the port 2 of a VNA (Keysight E5061B). Port 1 of the VNA is directly connected to the input of the DUT. Circuit stability is monitored with a spectrum analyzer connected at the output of the amplifier. A photograph of the setup is given in Fig. 5. As mentioned in the previous section, a crucial point of the technique relies on the negligible loading effect of the probe. We have measured the impedance of the active probe in our setup. The probe has an impedance of about 1100 Ω at 180 MHz that lowers down to 458 Ω at 500MHz. These figures are in good agreement with the probe characteristics provided in the datasheet [20].

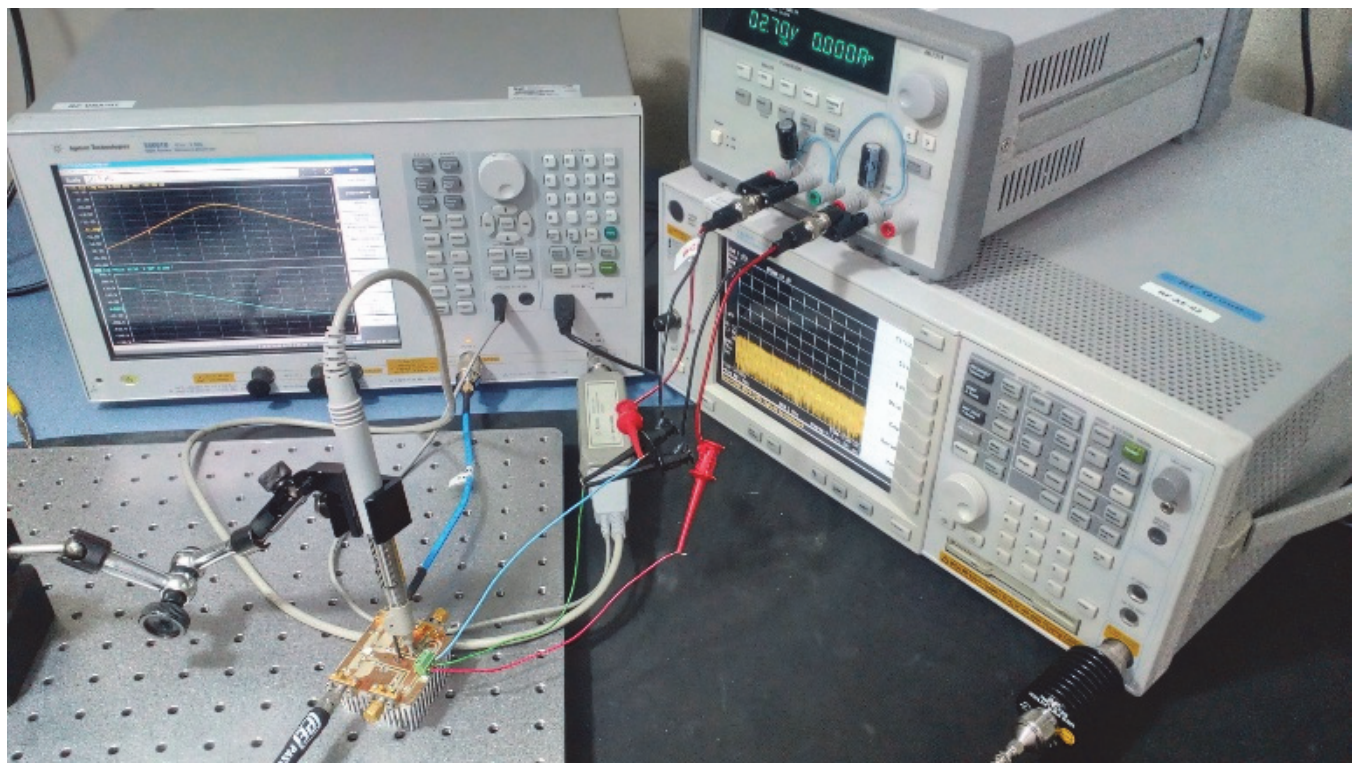

Fig. 5. Photograph of the measurement setup: port 1 of VNA injects small-signal $v_{gen}$ at the input of the amplifier and the active probe, connected to port 2 of VNA, senses at desired internal nodes; circuit stability is monitored with a spectrum analyzer connected at the output of the amplifier.

The prototype used for validation is a medium power L-band amplifier (Fig. 6) based on a single GaAs FET (FLU17XM). A schematic of the amplifier is depicted in Fig. 7. This prototype has been selected because it has additional RF observation ports (ports G and D in Fig. 6) connected to gate and drain bias paths in order to have access to the low frequency dynamics. The presence of these extra ports is essential for the validation of the technique because it allows the comparison of the poles obtained through in-circuit measurements with the poles extracted from the connectorized solution presented in [8]. The amplifier is mounted on an Arlon substrate. It presents a spurious oscillation at low frequency (about 190 MHz) for class AB bias when drain bias voltage $V_{DD}$ is higher than 3 V (Fig. 8). This suggests that, for $V_{DD}$ values lower than 3 V, a pair of complex conjugate poles will be very close to the RHP. Monitoring these low-damping poles is the goal of the proposed in-circuit characterization process.

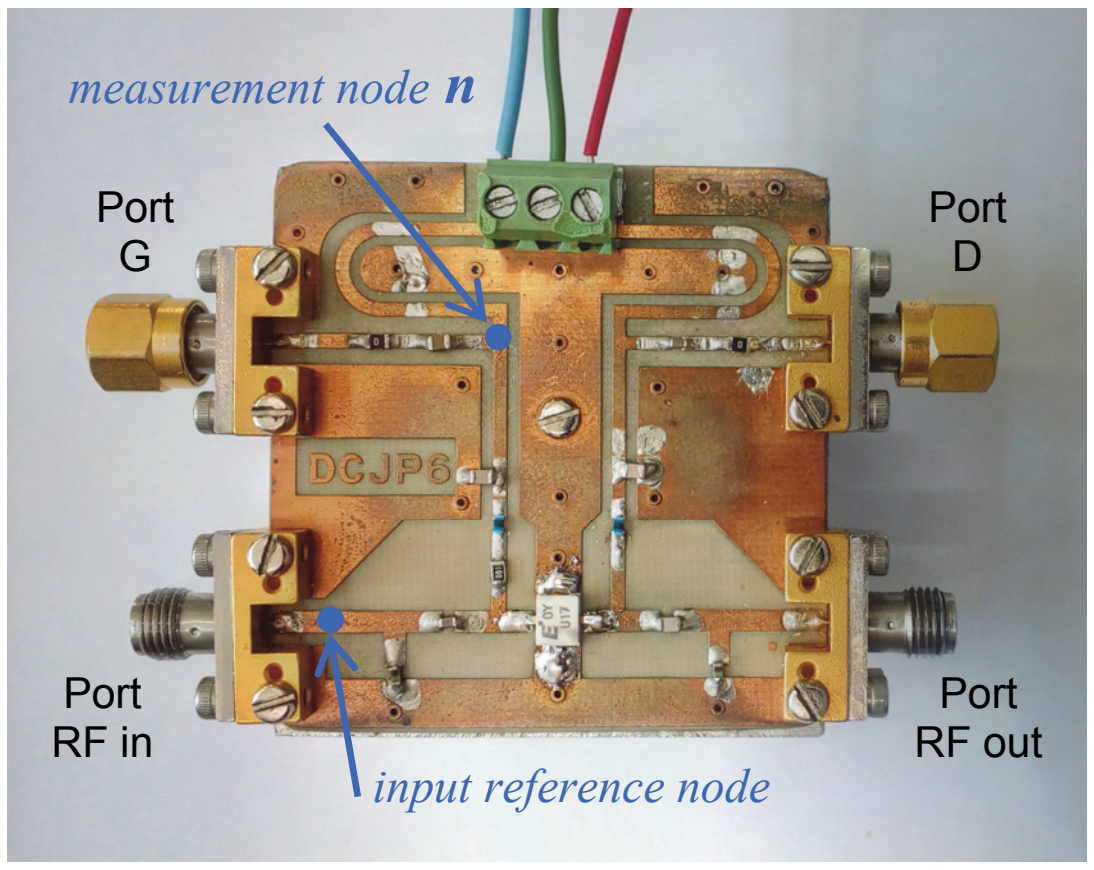

Fig. 6. Photograph of the medium power L-band amplifier based on a single GaAs FET, with additional observation RF ports G and D connected to gate and drain bias paths.

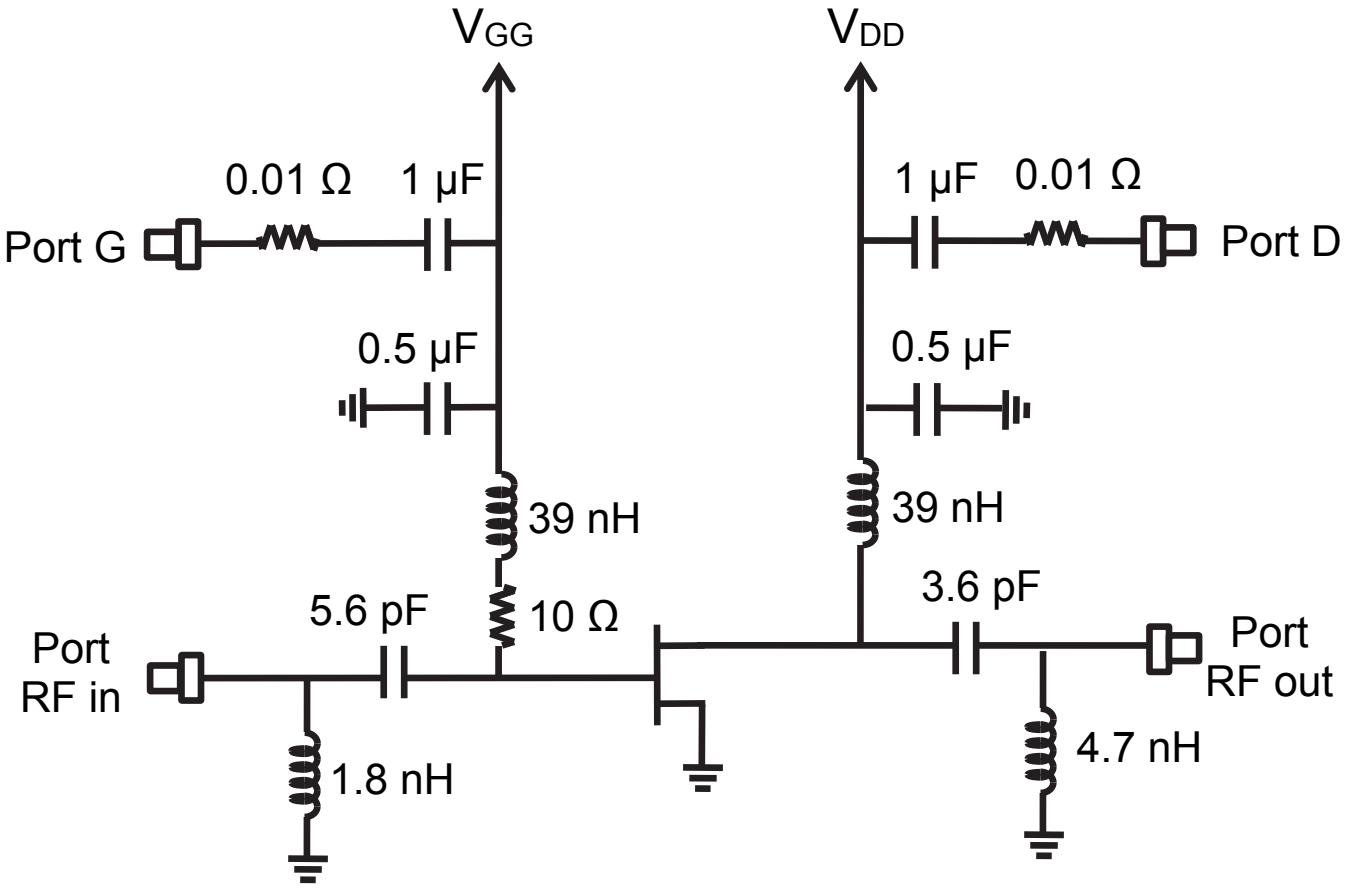

Fig. 7. Schematic of the medium power L-band amplifier based on a single GaAs FET (Fig. 6).

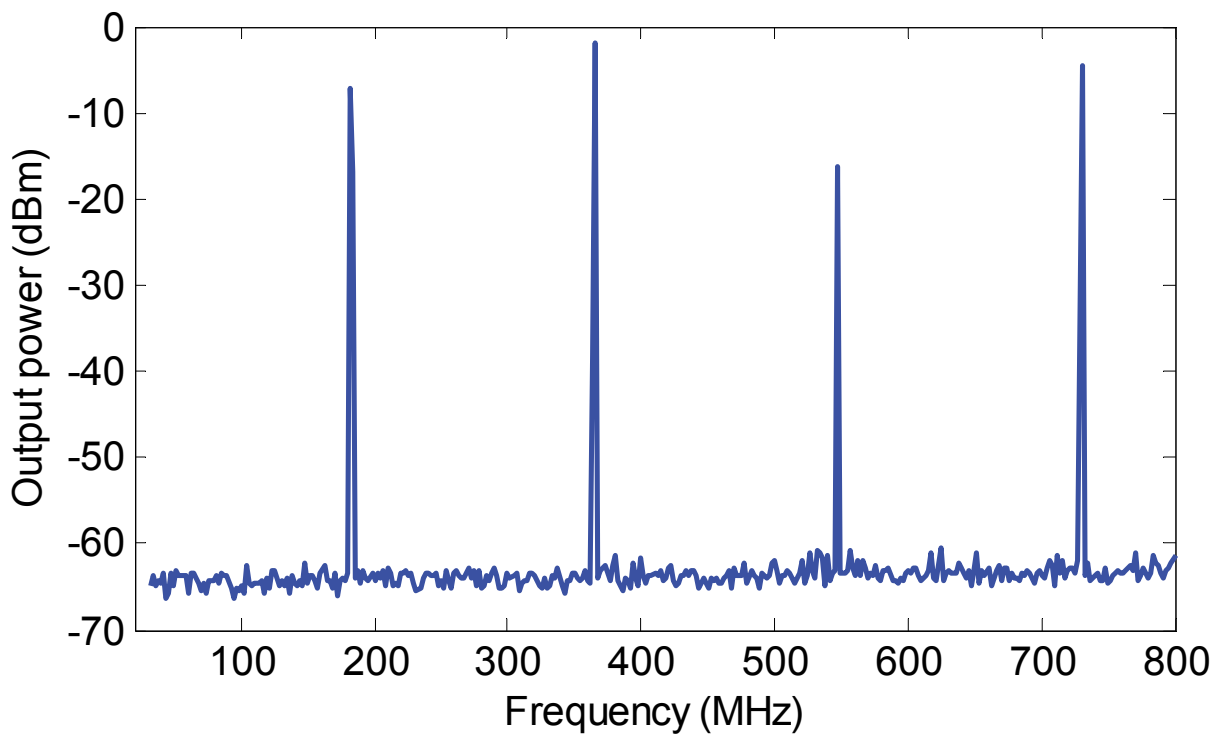

Fig. 8. Output spectrum of the L-band amplifier of Fig. 6 measured with $V_{GG}$= -2.1 V, $V_{DD}$ = 3.1 V. A low frequency spurious oscillation at about 190 MHz is observed.

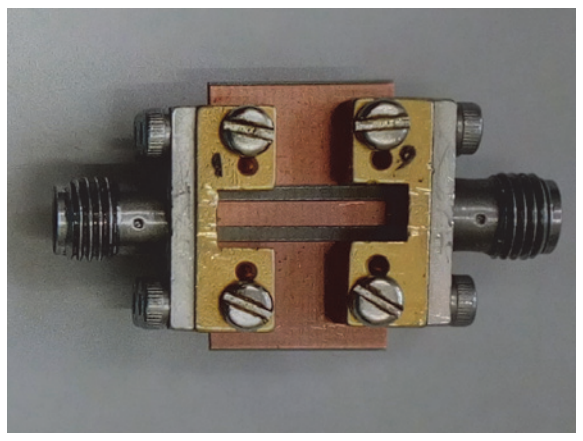

Fig. 9. Photograph of specific prototype to obtain S-parameters of input-block for de-embedding, fabricated on same substrate and with identical connectors to those of the amplifier under test.

Input reference node and internal measurement node *n* are marked on Fig. 6. The node *n* is selected on the gate bias network. In order to accurately de-embed the input block (the section from the input connector to input reference node indicated in Fig. 3), the specific prototype of Fig. 9 has been fabricated and measured. It has the same substrate, and identical connectors and input transmission line as the amplifier under study. Scattering parameters of the input block have been determined following the procedure explained in Section II.B.

### B. *Validation of the in-circuit measurements*

Following the methodology described in Section II, the closed-loop frequency response $H_{vn} = v_n / v_{gen}$ has been measured in-circuit for different values of $V_{DD}$ and fixed gate bias voltage $V_{GG}$ = -2.1 V (corresponding to a class AB bias). The magnitude and phase of the measured $H_{vn}$ frequency responses are plotted in Fig. 10. We can clearly observe a resonant behavior at about 190 MHz that sharpens as $V_{DD}$ increases, already indicating a reduction of the stability margin with poles getting closer to the RHP. This is confirmed by applying pole-zero identification to the responses of Fig. 10. Fig. 11 shows the pole evolution versus $V_{DD}$ corresponding to this critical resonance. Only positive frequencies of the complex plane are plotted for the sake of simplicity.

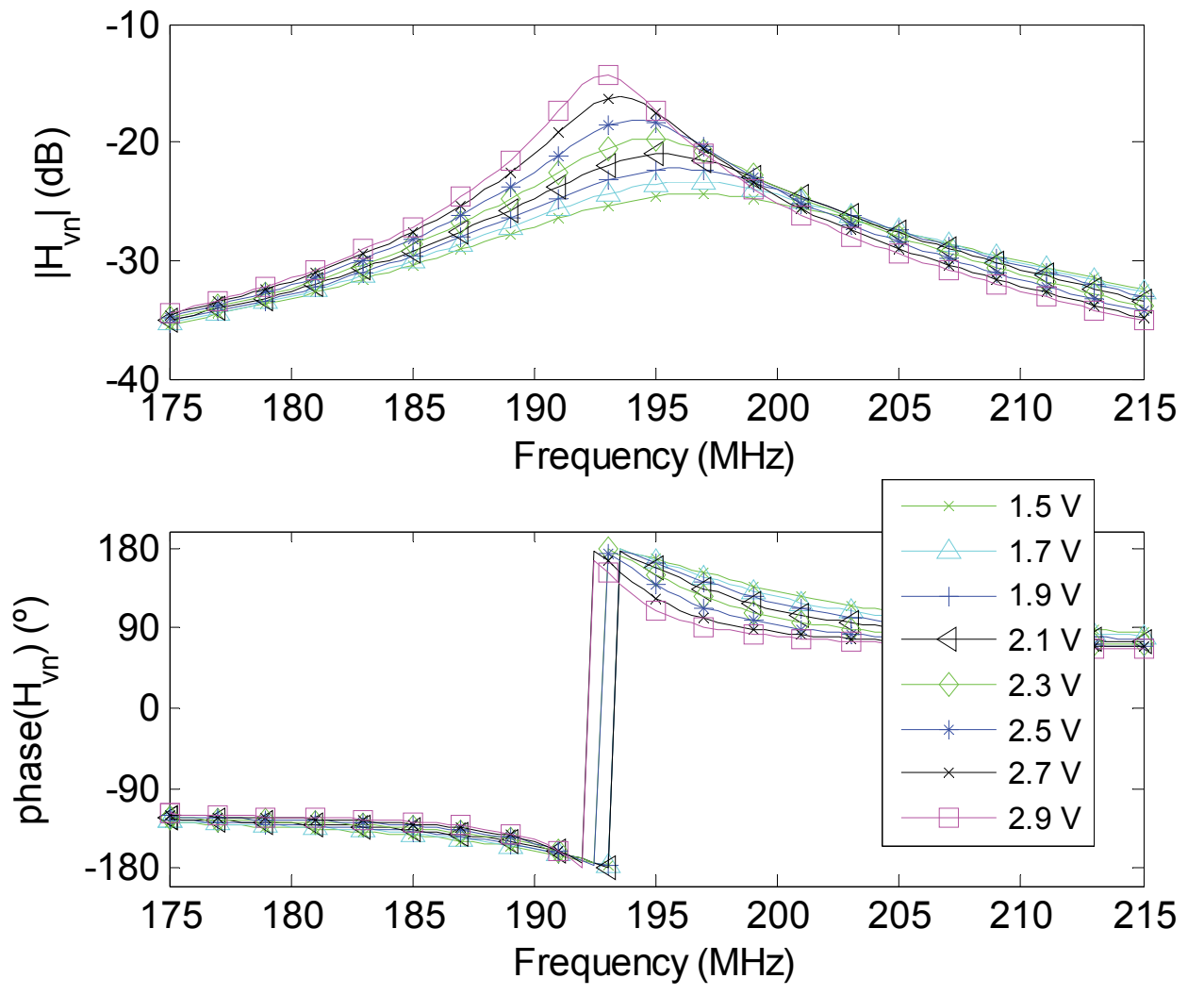

Fig. 10. In-circuit measurements: Magnitude and phase of closed-loop frequency responses $H_{vn}$ varying $V_{DD}$ from 1.5 to 2.9 V with $V_{GG}$ = -2.1 V. A sharpening resonance is clearly observed at about 190 MHz.

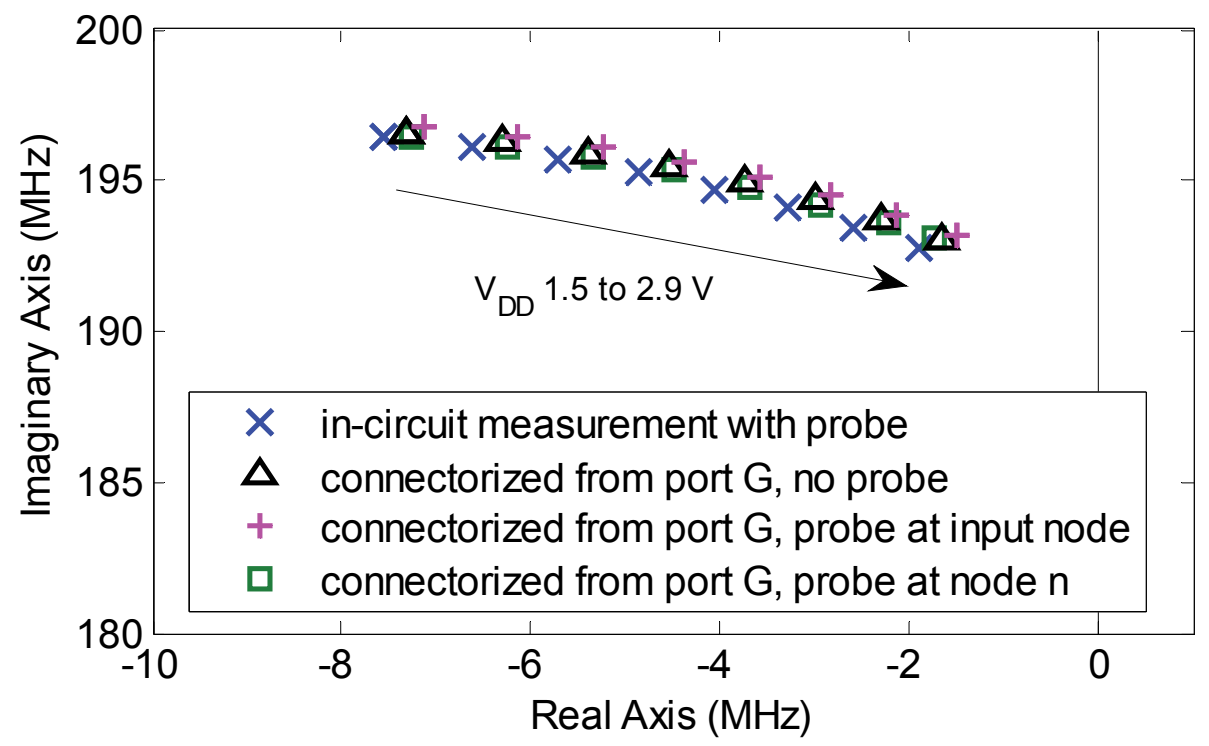

Fig. 11. Evolution of the critical poles of the amplifier varying $V_{DD}$ from 1.5 to 2.9 V with $V_{GG}$= -2.1 V: (×) obtained from identification of the in-circuit measured frequency responses $H_{vn}$ of Fig. 10. Superimposed are the evolutions obtained from the connectorized approach, measuring reflection coefficient at port G with no probe (Δ), with probe at input reference node (+) and with probe at node $n$ (□). Only positive frequencies of the complex plane are plotted.

In order to validate these measurements, we take advantage of the presence of the extra ports inserted in the bias paths (labeled G and D in Fig. 6), which give access to the low-frequency dynamics of the amplifier. In this way, we can use the connectorized approach proposed in [8] to get the low-frequency poles from a reflection coefficient measurement at port G. This procedure has been applied to the amplifier prototype for the same $V_{GG}$ and variable $V_{DD}$ bias values. The obtained pole evolution with this connectorized method is superimposed in Fig. 11, showing a very good agreement with the poles extracted from the in-circuit characterization.

To verify the negligible loading effect of the probe in the measurement nodes, connectorized measurements from port G are performed with the probe posed at input reference node and at measurement node $n$. The obtained poles have been superimposed in Fig. 11. Note that identical results are obtained with the probe posed at input reference node, at node $n$ or without the probe. The loading effect of the probe is thus negligible at the characterization frequencies.

## IV. Application to a Multistage Power Amplifier

The presence of additional observation ports on the bias lines is very useful to access the low-frequency dynamics and it served here to validate the in-circuit measurement approach. However, in most cases we will not have those extra ports available in the prototypes under study, especially in multistage amplifiers, in which their presence would be extremely cumbersome. In that case, the in-circuit characterization proposed here offers a solution to get the internal critical poles that may affect the amplifier instantaneous bandwidth. In this section, we show how the in-circuit characterization can be applied to a multistage amplifier to get the low-frequency critical poles and to determine which stage is responsible for the undesired dynamics.

The DUT is a hybrid L-band three-stage amplifier built in microstrip technology (Fig. 12). It is based on the same GaAs FET transistors (FLU17XM) as in the previous example. No extra ports for the observation of the low-frequency dynamics have been built-in. This amplifier presents different instabilities depending on several parameters [21]. A schematic of the amplifier is depicted in Fig. 13.

The in-circuit characterization process has been carried out at three internal nodes, one per stage, labeled $n_1$, $n_2$ and $n_3$ in Fig. 13. The same input reference node is considered for the three cases. The three closed-loop frequency responses $H_{vn1}$, $H_{vn2}$, $H_{vn3}$, have been measured for a constant $V_{GG}$ = -2.1 V and increasing $V_{DD}$ from 1 to 15 V. Their magnitude is plotted in Fig. 14. We can observe a clear resonant peak at about 130 MHz in $H_{vn2}$, while it is notably attenuated in $H_{vn3}$ and hardly noticeable in $H_{vn1}$.

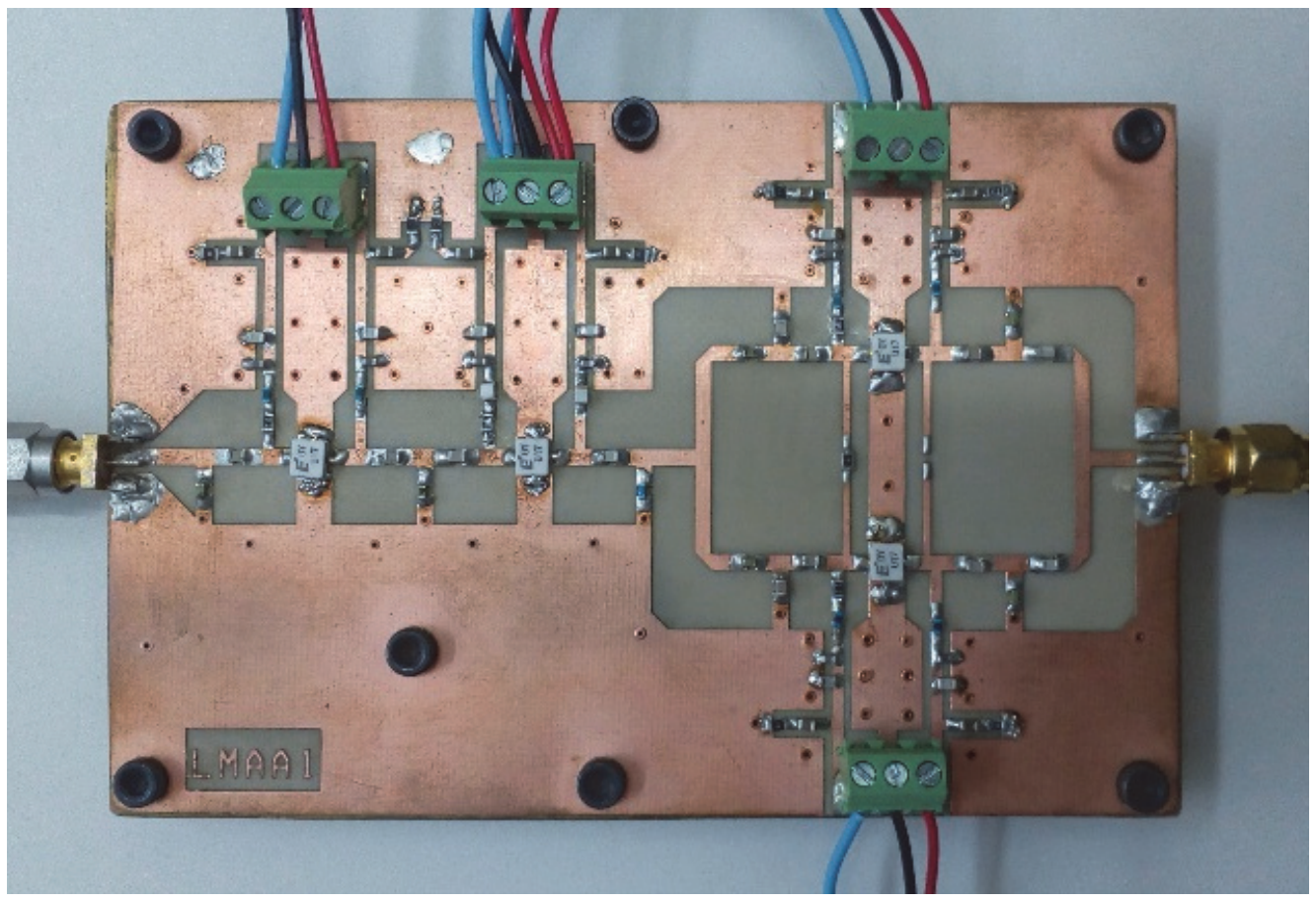

Fig. 12. Photograph of GaAs FET based L-band three-stage amplifier in microstrip technology. The amplifier presents different instabilities depending on several parameters.

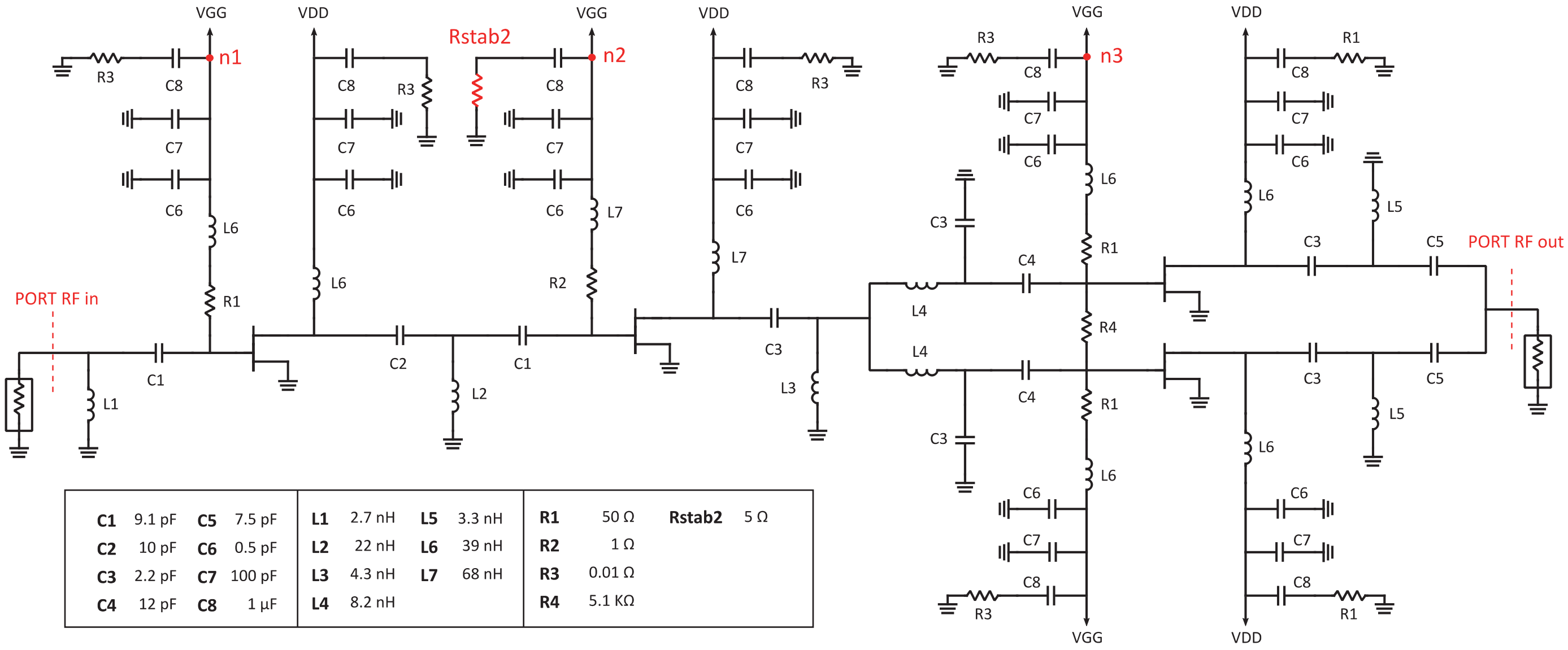

| | | | | | | | | | | | |
|---|---|---|---|---|---|---|---|---|---|---|---|
| C1 | 9.1 pF | C5 | 7.5 pF | L1 | 2.7 nH | L5 | 3.3 nH | R1 | 50 Ω | Rstab2 | 5 Ω |
| C2 | 10 pF | C6 | 0.5 pF | L2 | 22 nH | L6 | 39 nH | R2 | 1 Ω | | |
| C3 | 2.2 pF | C7 | 100 pF | L3 | 4.3 nH | L7 | 68 nH | R3 | 0.01 Ω | | |
| C4 | 12 pF | C8 | 1 µF | L4 | 8.2 nH | | | R4 | 5.1 KΩ | | |

Fig. 13. Schematic of the L-band three-stage amplifier of Fig. 12.

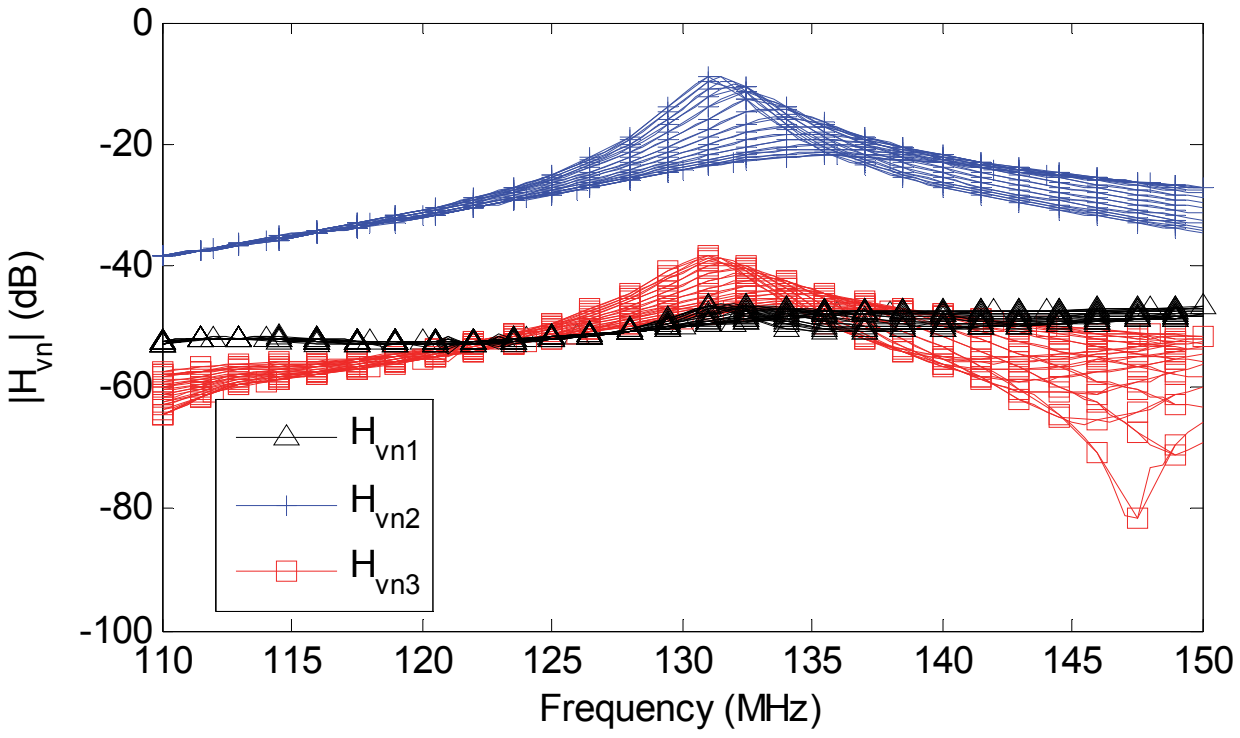

Fig. 14. Magnitude of measured closed-loop frequency responses $H_{vn1}$, $H_{vn2}$ and $H_{vn3}$ with $V_{GG}$ = -2.1 V and varying $V_{DD}$ from 1 to 15 V. A clear resonant peak at about 130 MHz is clearly observed in $H_{vn2}$, while it is notably attenuated in $H_{vn3}$ and hardly noticeable in $H_{vn1}$.

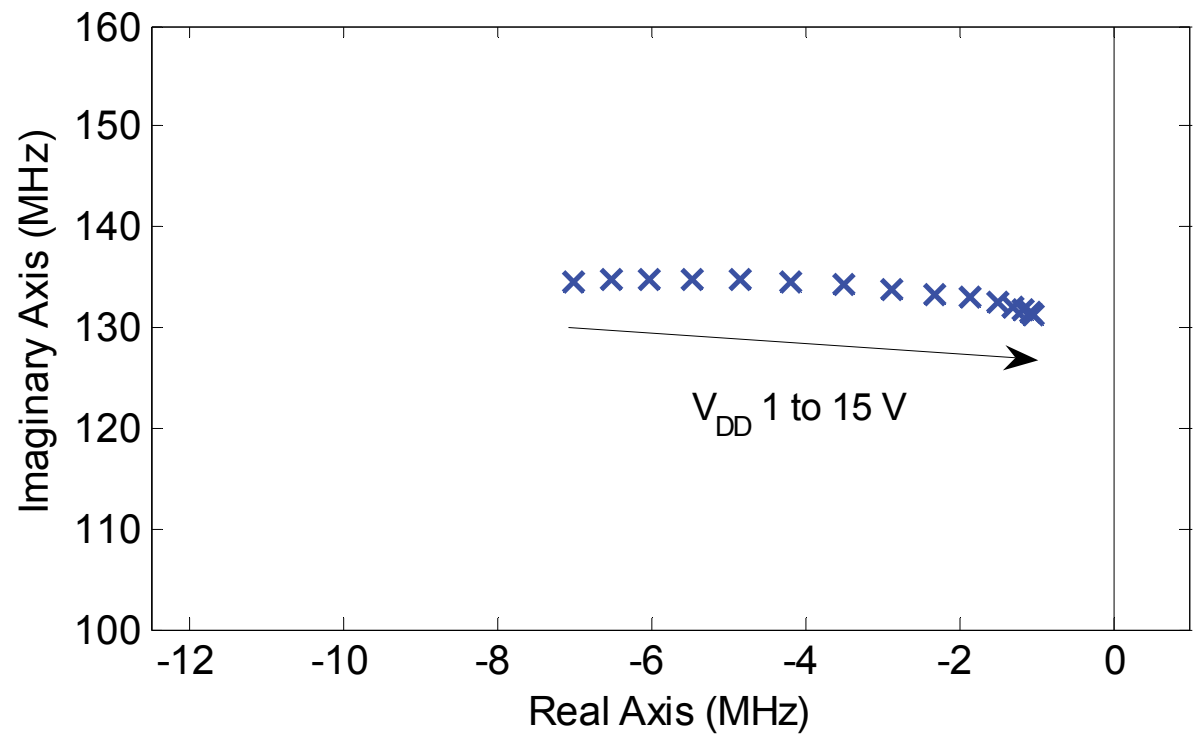

Fig. 15. Evolution, approaching RHP, of the critical poles of the amplifier with $V_{DD}$ 1 to 15 V, obtained from identification of in-circuit measurement of $H_{vn2}$ ($V_{GG}$ = -2.1 V). Only positive frequencies of the complex plane are plotted.

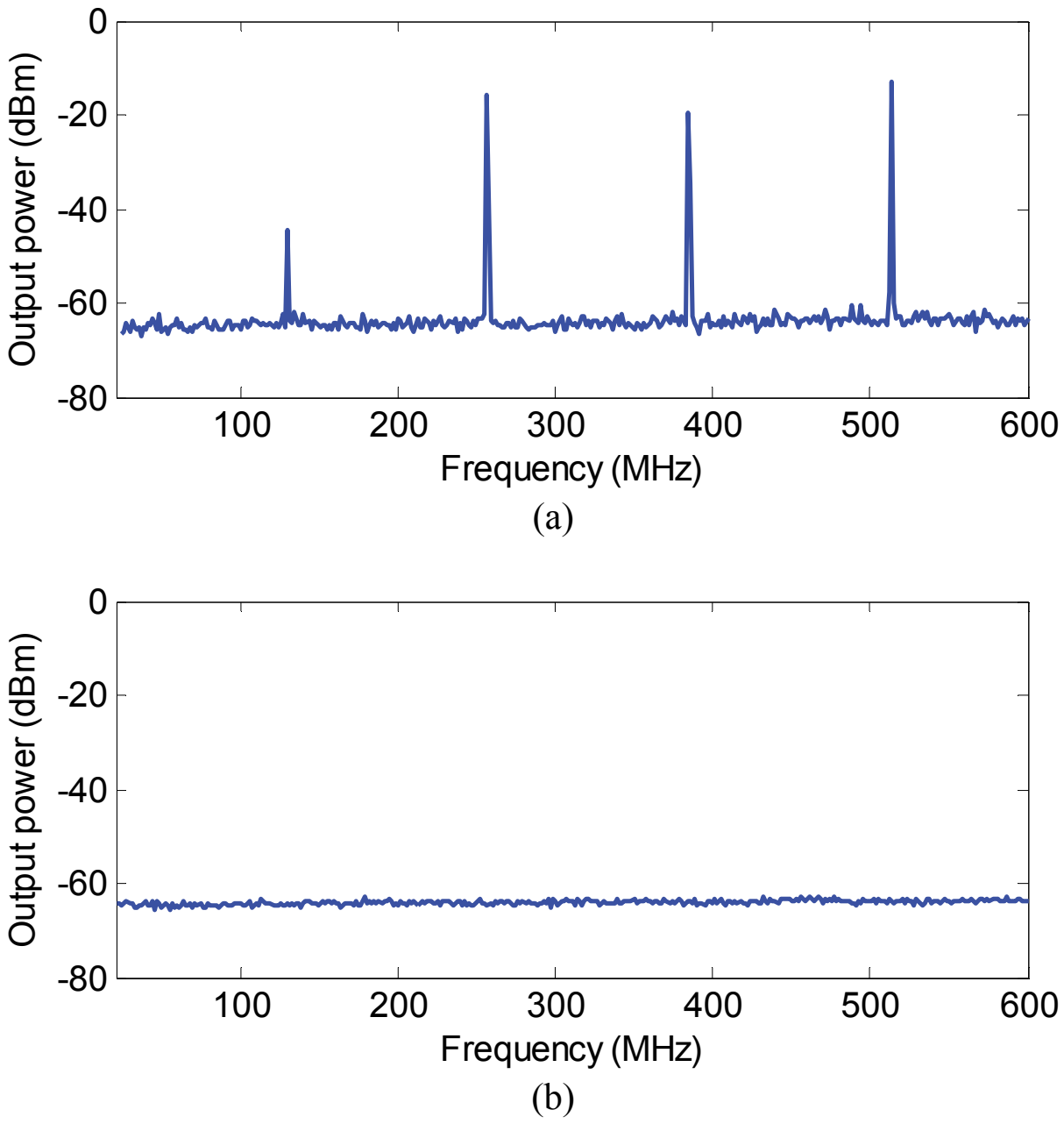

Fig. 16. Output spectrum of the amplifier, measured with $V_{GG}$ = -2 V and $V_{DD}$ = 5 V, (a) original version of the amplifier showing spurious oscillation at about 130 MHz. (b) stabilized version of the amplifier after increasing stabilization resistance of the second stage ($R_{stab2}$ in Fig. 13) from 5 to 20 Ω.

Applying pole-zero identification to $H_{vn2}$, a critical pair of complex conjugate poles that shift rightwards in the LHP is detected (Fig. 15). These poles, although stable, approximate dangerously to the RHP and may easily become unstable with slight changes in the circuit conditions such as temperature or bias. Actually, changing gate bias from $V_{GG}$ = -2.1 V to $V_{GG}$ = -2 V a spurious oscillation appears at about 130 MHz for drain bias $V_{DD}$ = 5 V, as shown in Fig. 16(a). With $V_{GG}$ = -2 V and varying $V_{DD}$, the frequency response $H_{vn2}$ has been measured and identified, obtaining the pole evolution depicted in blue in Fig. 17. With this bias, the critical poles shift rightwards comparing to the ones shown in Fig. 15, as it can be seen in the pole trajectory nearest to the imaginary axis. Four stable poles can be seen now, corresponding to $V_{DD}$ values from 1 to 4 V, in good agreement with the oscillation for $V_{DD}$ = 5 V shown in Fig. 16(a).

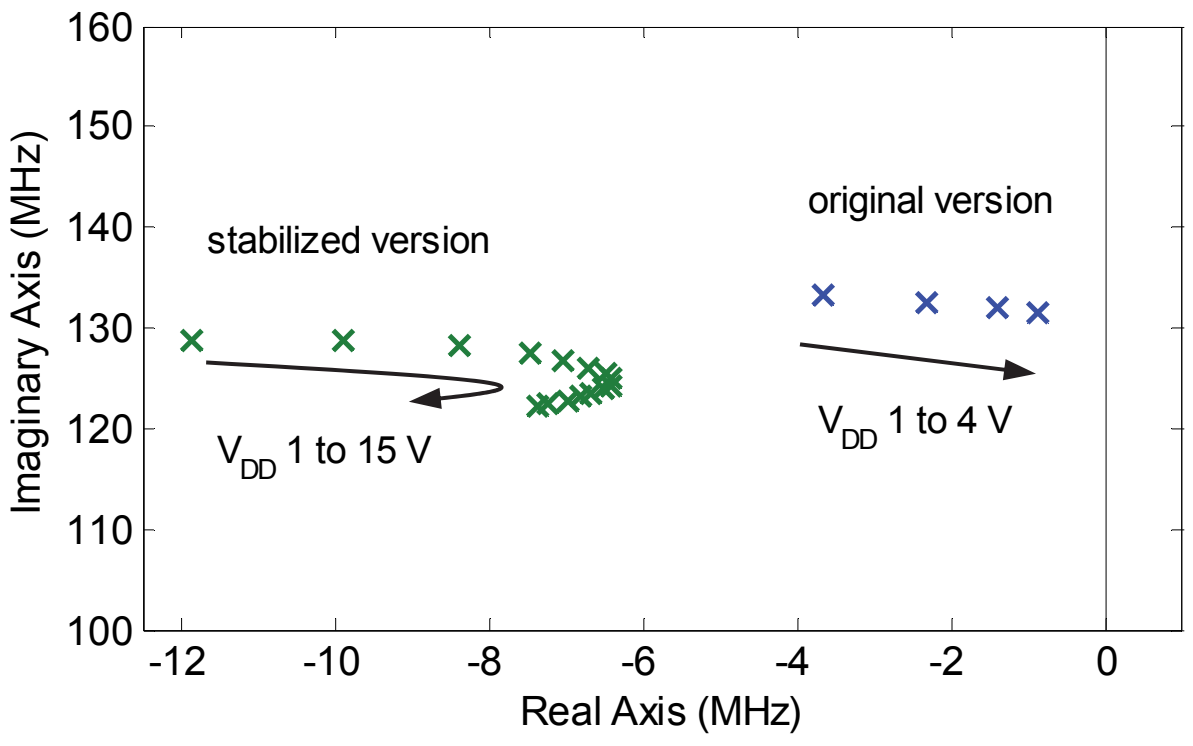

Fig. 17. Evolution of the critical poles of the amplifier with $V_{DD}$ obtained from identification of in-circuit measurement of $H_{vn2}$ ($V_{GG}$ = -2 V) for the original version of the amplifier ($R_{stab2}$=5 Ω) and for the stabilized version ($R_{stab2}$=20 Ω).

In order to improve the stability margin, it is important to investigate which is the stage responsible for this critical resonance. Previous techniques to determine the origin of the critical poles [17], [18], [21] are not directly applicable to voltage transfer functions like $H_{vn1}$, $H_{vn2}$ and $H_{vn3}$. Actually, in a multi-stage power amplifier, a critical resonance associated with a particular stage may be clearly noticeable in voltage transfer functions obtained at next stages. The resonance will be attenuated or amplified according to the gain and the inter-stage matching at the resonant frequency. On the contrary, due to the unilateral nature of the transistor, the resonant behavior will be barely perceptible in voltage transfer functions obtained in previous stages. This is actually what is happening in the multi-stage amplifier under study, where the clear resonant peak at about 130 MHz in $H_{vn2}$ is attenuated in $H_{vn3}$ and almost unnoticeable in $H_{vn1}$ (Fig. 14). As a general rule, the origin of the critical resonance will be the first stage in which the resonance is clearly observable. In the case of the circuit under analysis, this is the second stage of the amplifier.

We have applied the simulation techniques in [17], [18], and [21] to the three-stage amplifier in order to confirm the location of the critical resonance. Three admittance type frequency responses $Y_1$, $Y_2$, $Y_3$ have been simulated for $V_{GG}$ = -2.1 V and $V_{DD}$ = 7 V. For that, small-signal voltage sources are sequentially injected in series at the gate bias line of each stage [17], [18]. The frequency response is calculated as the ratio between the current flowing through the branch to the voltage of the small-signal source [17], [18]. Magnitude and phase of $Y_1$, $Y_2$ and $Y_3$ are plotted in Fig. 18. A Multiple-Input Multiple-Output identification [21] is applied to the set of three simulated frequency responses. The obtained pair of complex conjugate poles is shown in Fig. 19. Superimposed in Fig. 19 is the pair of complex conjugate poles obtained experimentally from $H_{vn2}$ for the same bias condition. A very good agreement is obtained between the simulated and the measured poles.

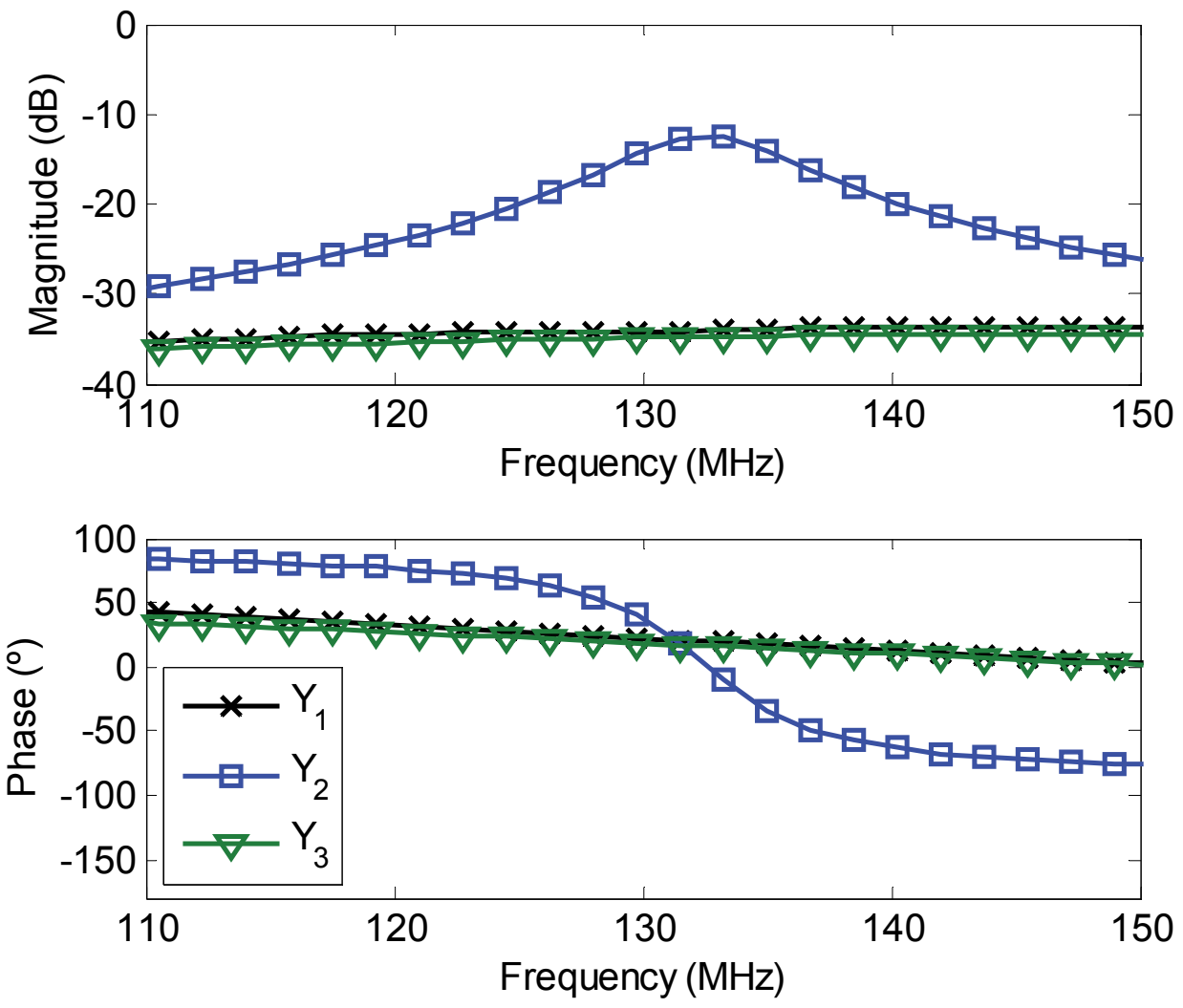

Fig. 18. Magnitude and phase of admittance type frequency responses $Y_1$, $Y_2$, $Y_3$ simulated for $V_{GG}$ = -2.1 V and $V_{DD}$ = 7 V. The critical resonance is only clearly noticeable at the admittance response $Y_2$ obtained at the second stage.

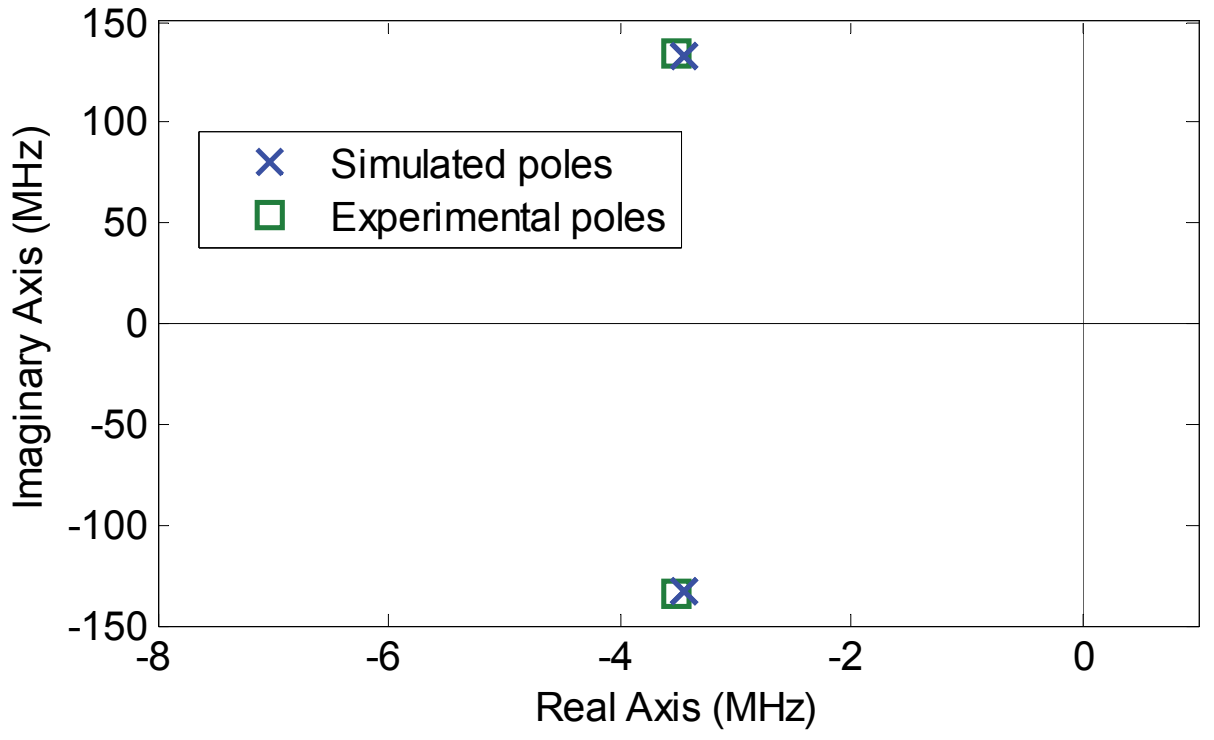

Fig. 19. Multiple-Input Multiple-Output identification results of simulated $Y_1$, $Y_2$, $Y_3$ frequency responses of Fig. 18 (×). Critical poles of the amplifier obtained experimentally from $H_{vn2}$ for $V_{GG}$ = -2.1 V and $V_{DD}$ = 7 V (□).

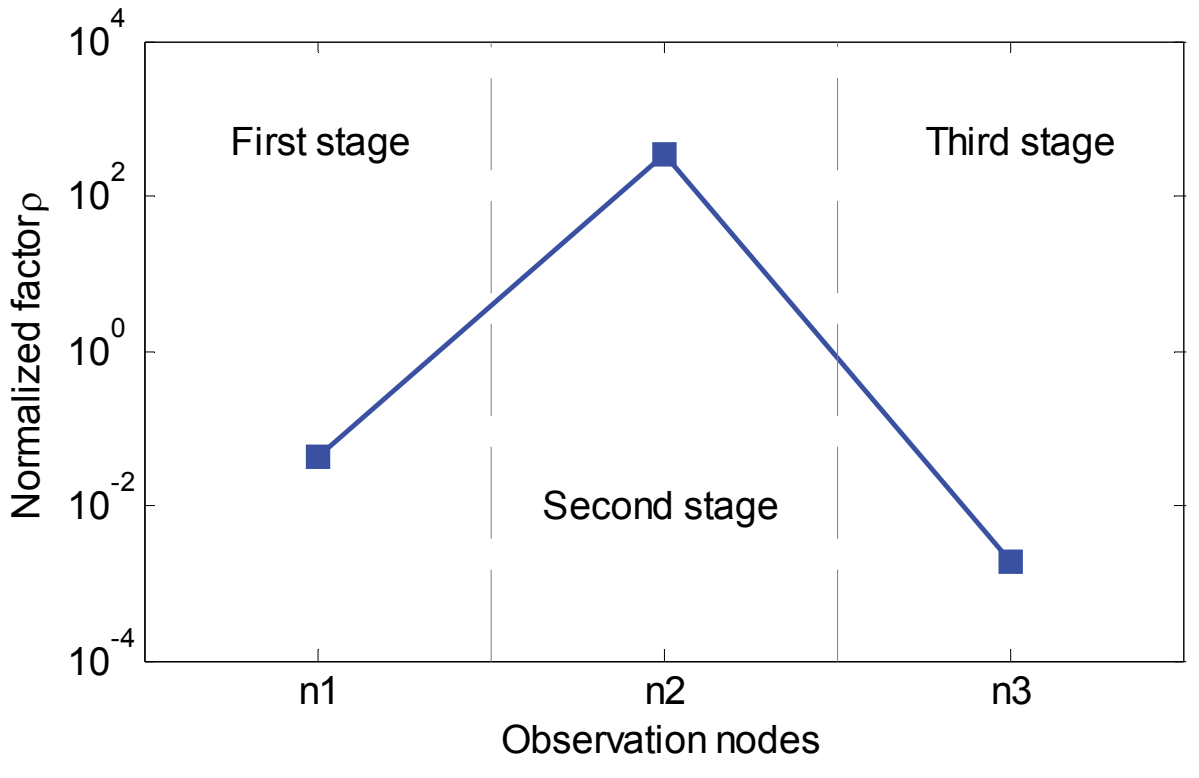

Fig. 20. Residue analysis of simulated critical poles of Fig. 19. Normalized residue at second stage is significantly larger, indicating that the origin of the unstable dynamics is in the second stage.

We can observe in Fig. 18 that the critical resonance is only clearly noticeable at the second stage. From [17], [18] this suggests that the origin of the resonance is in the second stage, as deduced before. For a quantitative confirmation, we have applied a residue analysis [21] to the Multiple-Input Multiple-Output identification of the three frequency responses $Y_1$, $Y_2$ and $Y_3$. Results are shown in Fig. 20. For the critical pair of complex conjugate poles, the normalized residue obtained at the second stage is much larger than those of first and third stages. This implies that the origin of the unstable dynamics is in the second stage undoubtedly.

Once the origin of the instability has been detected to be in the second stage, a proper stabilization solution can be straightforwardly chosen and applied. In our case, the resistance $R_{stab2}$ of the stabilization RC network of the second stage, labeled in Fig. 13, has been increased from 5 to 20 Ω. In order to verify the stabilization solution, the output spectrum has been measured again for $V_{GG}$ = -2 V and $V_{DD}$ = 5 V. As it can be seen in Fig. 16(b), the spurious oscillation shown in Fig. 16(a) has disappeared. The critical poles have been measured again with $V_{GG}$ = -2 V and $V_{DD}$ variable from 1 to 15 V. The obtained evolution is depicted (in green) in Fig. 17, where it is compared to the evolution corresponding to the original version (in blue). As it can be seen the critical poles have significantly moved leftwards, providing a better stability margin and thus making the amplifier more robust against slight changes in the operating conditions. Eventually, it has also been verified experimentally that stabilization resistors in first and third stages have no effect on circuit stabilization in this prototype.

## V. Extension to Large-Signal Steady State Regimes Forced by the Input Power

In the previous sections only dc or small-signal case has been addressed. However, measuring the stability margin associated with the low frequency resonance as the input drive increases is very important for high-power amplifier (HPA) design and diagnosis. Analogously to [22], the approach needs substantial modifications from the previous small-signal analysis because of the Periodic Linear Time Variant (PLTV) nature of the system resulting from the linearization of a periodic large-signal regime. The experimental implementation is more delicate than the *dc* case because the required non-perturbation condition of the probe becomes more challenging. On the one hand, the impedance of the probe must be significantly larger than the impedance of the observation node at the low frequencies of the small-signal characterization, like in the dc case. On the other hand, the probe impedance must also be larger than the node's impedance at the RF operating frequency $f_0$ and its harmonics in order to avoid the modification of the periodic large-signal steady state. We explain in the following the main modifications of the methodology to adapt to large-signal regimes.

As in [22], a power combiner at the input of the DUT is added to the setup (Fig. 21). The low frequency small-signal from port 1 of the VNA is connected to one of the combiner's ports while the large-signal drive at $f_0$ is injected to the other. This combiner needs to be broadband to handle the low frequency and the RF frequency. In our case a resistive power combiner is used because power losses are not a concern in the setup. Given that isolation of resistive combiners is not high, a VNA with a large linear dynamic range is needed to ensure that input power $P_{in}$ at $f_0$ does not saturate the VNA receiver at the low frequencies of the measurement $f_{test}$. This can be verified by measuring, through the power combiner and for different values of $P_{in}$, the scattering parameters of a passive device as explained in [22].

The most important challenge to overcome in this case is to guarantee the non-perturbing condition of the probe in large-signal conditions. At the high operating frequency $f_0$ and harmonics $nf_0$, the probe does not present "high-impedance" anymore. As a result, its connection at a particular node of the circuit (either node $n$ or the input reference node) can, in principle, drastically modify the large-signal steady state, which would invalidate the approach.

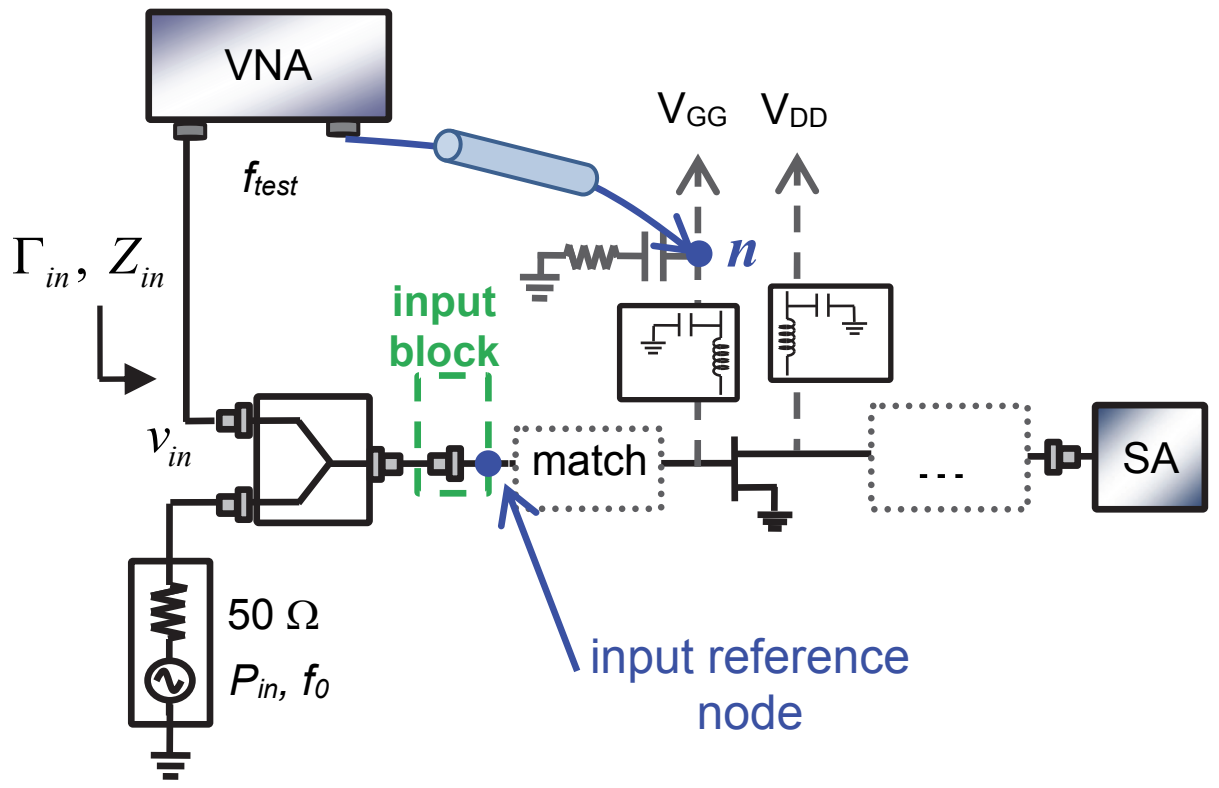

Fig. 21. Schematic of measurement setup for large signal regime: RF large-signal drive with input power $P_{in}$ at $f_0$ and the small signal from port 1 of VNA at $f_{test}$ are combined at the input of the circuit. As before, the probe, connected to port 2 of VNA, senses at an internal node $n$.

In order to avoid the perturbation when measuring at the internal node $n$, we propose to use as sensing positions only nodes located on the bias paths once the RF has been decoupled. These locations are well-suited to observe the low-frequency dynamics, while at the same time, the filtering of the RF signal makes the loading effect of the probe negligible at $f_0$ and $nf_0$. This can be easily verified in a spectrum analyzer monitoring the spectrum at the output RF port with and without the probe connected at internal node $n$.

However, with the RF large-signal on, connecting the probe at the input reference node would have a big impact on the large-signal steady state. To avoid perturbing the steady state when determining the voltage transfer function $H_n$, a single $v_{ref}$ measurement with the probe at reference node is performed without the large-signal applied. Note that this implies measuring $v_n$ and $v_{ref}$ in different configurations. $v_n$ is measured with the input power applied while $v_{ref}$ is measured without applying the input power drive. However, there is no drawback in using two different measurement configurations for $v_n$ and $v_{ref}$ provided that $v_{gen}$ is the same for the two configurations. For consistency, the transfer function between $v_{gen}$ and $v_{ref}$, $H_{input}$ (5), has to be fully characterized in the same conditions than $v_{ref}$. Therefore, $\Gamma_{in}$ and $Z_{in}$ in (6) and (7) must be measured without the input power applied. Note also that $\Gamma_{in}$ and $Z_{in}$ are now the reflection coefficient and the impedance seen at the input port of the combiner that is connected to the VNA. In addition, the scattering parameters $S_{ij}$ in (7) correspond to the cascade of the power combiner with the input block. To obtain these, the scattering parameters of the power combiner relating the port connected to VNA to output port (Fig. 21) have to be characterized previously.

This extended methodology has been applied to the medium power L-band amplifier of Section III with an input signal at 1.2 GHz. This prototype has a spurious oscillation that starts for input power $P_{in}$ > 15.3 dBm, when biased at $V_{GG}$ = -1.7 V, $V_{DD}$ = 7 V. Fig. 22(a) shows the stable output spectrum for a $P_{in}$ of 14 dBm. Fig. 22(b) shows the output spectrum for a $P_{in}$ of 15.4 dBm, where a spurious oscillation at approximately 165 MHz is generating mixing products about $f_0$ and harmonics.

For the characterization with the high impedance probe, we have selected the same node *n* on the gate bias line that was used in the dc case of Section III (Fig. 6). To verify that the probe is not perturbing the large-signal regime, we have measured the output spectrum with the probe connected at node *n* for 14 dBm. This has been superimposed to the output spectrum measured without the probe connected in Fig. 22(a). The two spectra are identical, which confirms the non perturbation of the probe.

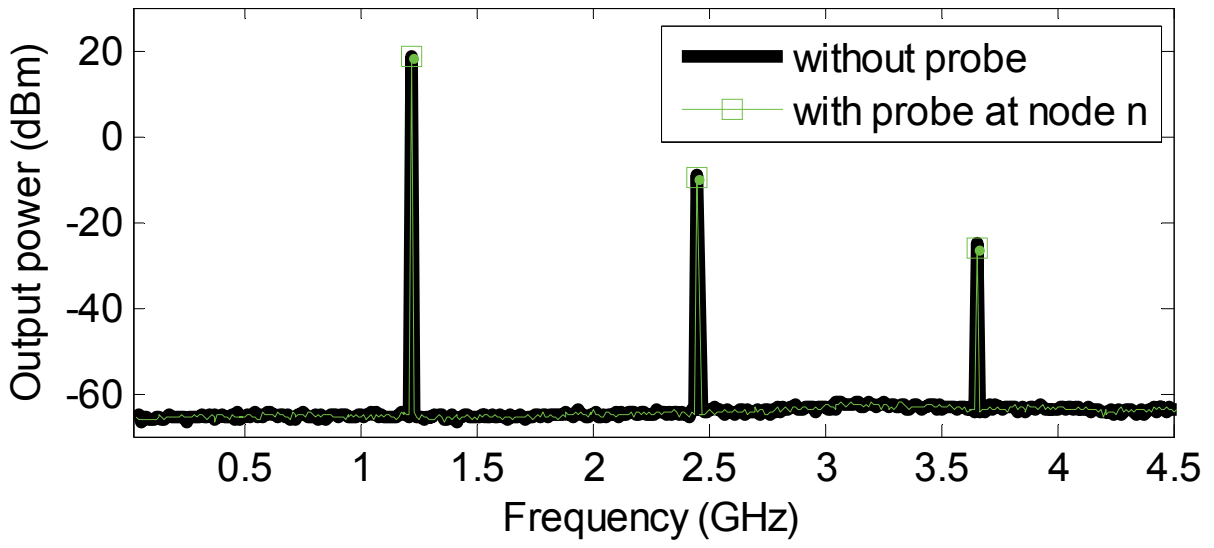

(a)

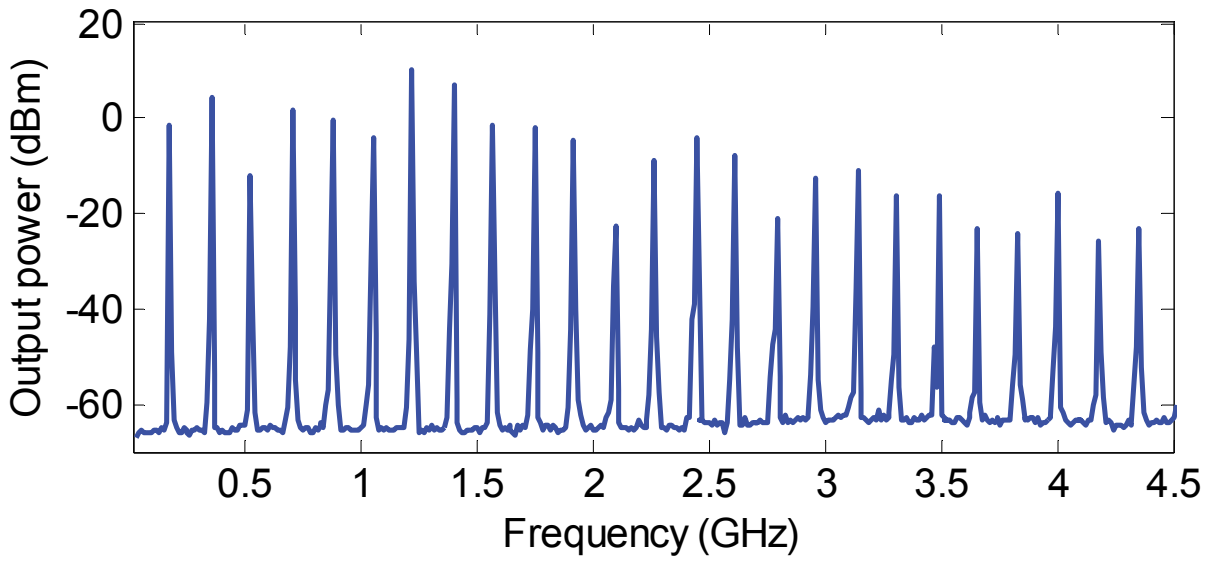

(b)

Fig. 22. Output spectrum of the single-stage L-band amplifier biased at $V_{GG}$ = -1.7 V, $V_{DD}$ = 7 V, with a pumping signal at 1.2 GHz: (a) Stable for $P_{in}$ 14 dBm; (b) Unstable for $P_{in}$ 15.4 dBm.

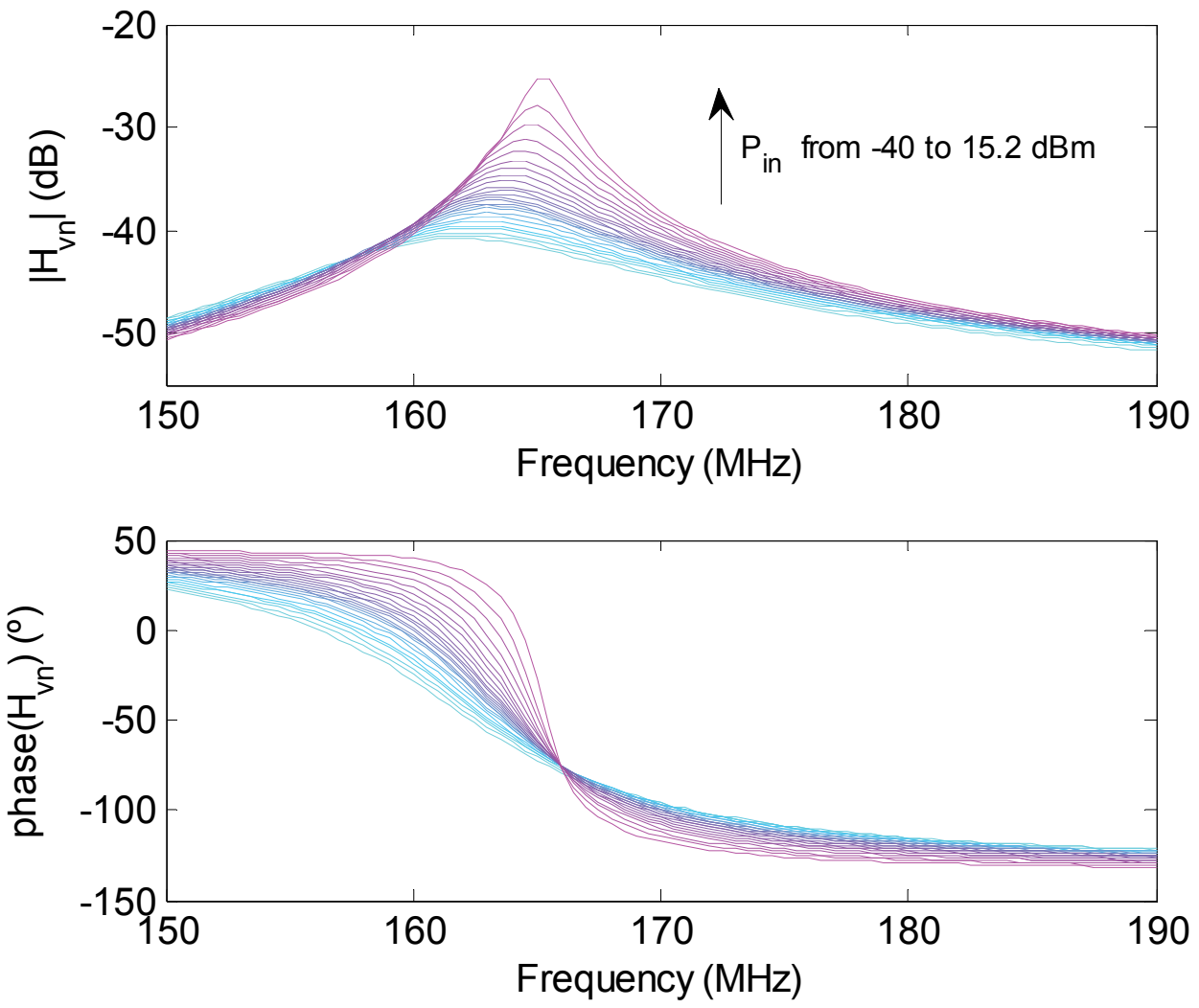

Fig. 23. Frequency responses $H_{vn}$ obtained from in-circuit measurements for $P_{in}$ varying from -40 to 15.2 dBm.

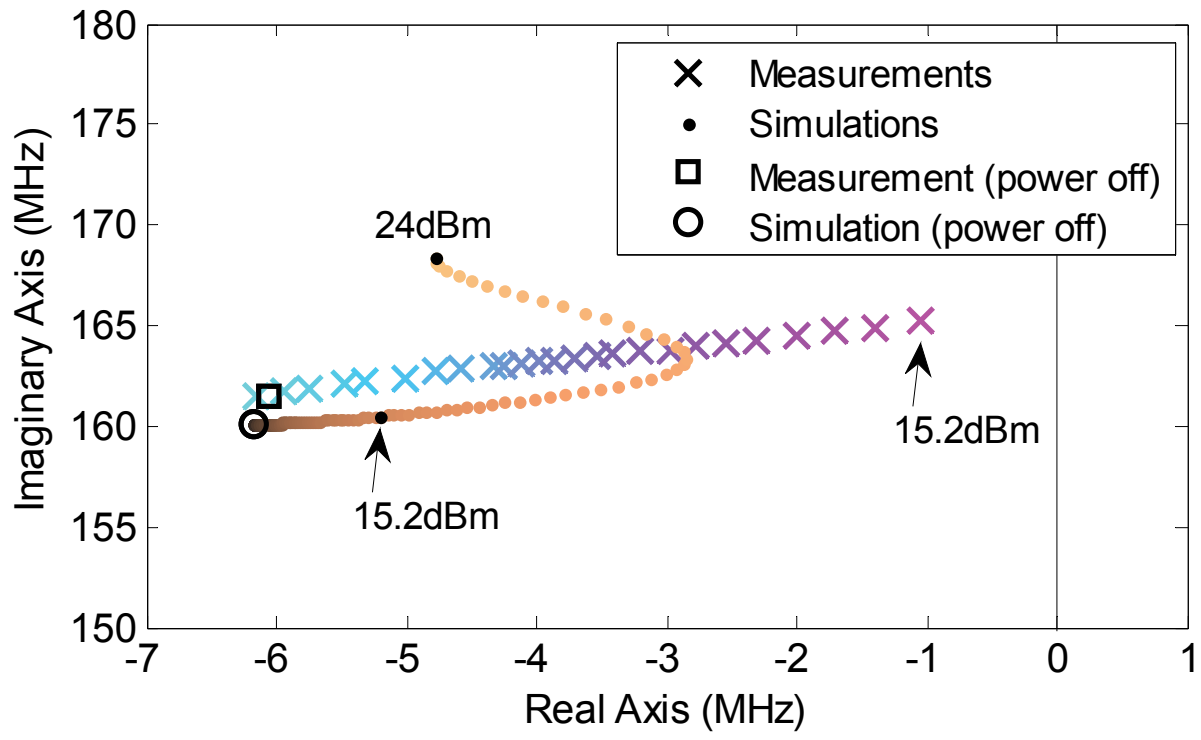

Fig. 24. Pole evolution for $P_{in}$ -40 to 15.2 dBm obtained from identification of frequency responses, both from in-circuit measurements and from simulation. In addition, the poles corresponding to dc case, i.e. no power applied, are superimposed.

The magnitude and phase of the frequency response $H_{vn}$ measured for varying $P_{in}$ is plotted in Fig. 23. The corresponding pole evolution after identification is shown in Fig. 24. The critical poles shift rightwards as $P_{in}$ is increased. For $P_{in}$ = 15.2 dBm poles are about to cross to the RHP, in agreement with the unstable behavior of the prototype measured at the spectrum analyzer for $P_{in}$ = 15.4 dBm. The pole obtained for dc case, i.e. without $P_{in}$, has also been depicted, showing a very good agreement with low $P_{in}$ values.

We have tried to duplicate these results by simulation. Unfortunately, the non-linear model of the device was not able to predict the pole evolution toward instability. The simulated poles are superimposed in Fig. 24. We can observe how they follow a different path than the measured ones, even turning more stable as $P_{in}$ is increased. They do not predict the instability that was experimentally confirmed. Actually, it is in these circumstances of model failure when an experimental approach to obtain the circuit poles may become most valuable.

## VI. Practical Limitations and Discussion

The frequency range of applicability of this technique is limited by the loading of the probe, which augments as frequency increases because its impedance lowers. For accurate measurements, the impedance of the probe must be significantly larger than the impedance of the sensing node. Therefore, as the frequency increases, fewer nodes are suitable for sensing: Only those nodes whose impedance is still significantly lower than the one of the probe. In order to quantify better this non-perturbing condition in the context of pole-zero identification, we have simulated the complete measurement setup including a model of the probe with varying impedance. For probe impedances ten times larger than the impedance of the node, the variations in the obtained critical poles are negligible for all the amplifiers we tested. Even though this ten-to-one ratio is not a general rule because it may depend on the particular circuit under study, it gives enough certainty on the fulfillment of the non-perturbation condition.

With our particular high-impedance probe and setup, critical poles up to 500 MHz could be measured with reasonable confidence. This limit is normally sufficient to analyze the low frequency dynamics of power amplifiers [23]. To reach higher frequencies, a practical limiting factor is the inductance introduced by the length of the ground wire that connects the probe to the board. This inductance resonates with the probe capacitance and lowers its impedance. To extend the frequency range we could either reduce the length of the ground lead as much as possible, restrict the sensing to nodes with very low impedance, or use a high-impedance probe with better performances (lower input capacitance).

Another limitation may come from using the RF input port to apply the input signal $v_{gen}$ because the circuit will be mismatched at the low frequencies of the characterization. In addition, the input signal might be attenuated as it flows through the stages reducing the sensitivity. However conventional VNAs are able to measure very low received signals thanks to their typical large dynamic range. In the event that the level of the received signal at a sensing node were too low, a noisy voltage transfer function $H_{vn}$ would be measured. This can be used as an indication of a very poor observability at that node. In this case, the output power at port 1 of the VNA can be increased so as to augment the received signal and improve the measurement. Circuit linearity with respect to that signal must be respected, obviously. In the examples we have tested we were able to accurately measure the resonances even though the amplifiers were highly mismatched at some of the characterization frequencies. We use for that an output power of -30 dBm at port 1 of the VNA in the dc steady state case. In the large-signal case, output power at port 1 was set to -20 dBm to compensate for the attenuation of the power combiner. Linearity of the results with respect to the small-signal excitation was verified as in [22].

Additionally, critical poles cannot be characterized arbitrarily close to the imaginary axis because circuit linearization is not valid in the immediate neighborhood of the bifurcation point (the value of the parameter for the onset of the oscillation) [4]. As a consequence, pole-zero identification may become problematic extremely close to the bifurcation, yielding inconsistent results. This limit cannot be quantified in general and may be different for each circuit. In the examples analyzed in this work, it has been possible to obtain the critical poles close enough to the bifurcation in order to get parametric pole evolutions that are consistent with the observed onset of the oscillations.

Although the proof of concept of the methodology has been carried out on hybrid prototypes, it could also be extended to the characterization of millimeter-wave integrated circuits (MMIC). For that purpose, high-impedance probes specifically designed for on-wafer characterization could be used as in [24] to probe at predefined pads. In that case, the small-signal would be applied through a 50 Ω ground-signal-ground (GSG) probe at the input of the circuit. The effect of this probe has to be accounted for in the de-embedding process of the input block. For that, a 2-port S-parameter block characterizing the GSG probe has to be determined as in [24].

## VII. Conclusion

An experimental methodology has been proposed for in-circuit characterization of low-frequency stability margins in power amplifiers. The methodology makes use of a high impedance probe to obtain calibrated closed loop frequency responses at internal circuit nodes. In order to validate the technique, results have been firstly compared to a connectorized approach on a specifically designed prototype. The methodology is applied to the detection and fixing of low-frequency critical resonances in a three-stage amplifier to increase its usefulness. Eventually, the methodology has been adapted to analyze the stability margins in large-signal conditions.

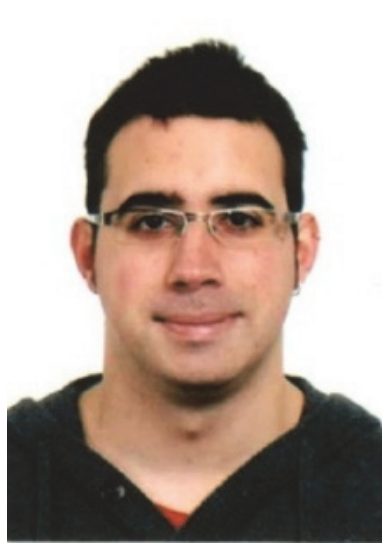

**Jose Manuel Gonzalez** received the M.Sc. degree from ENSEEIHT, Toulouse, France, in 2008, and the Ph.D. degree in high-frequency electronics from the Université de Limoges, France, and Carleton University, Ottawa, Canada, in 2011. In 2012 he joined KAUST (Saudi Arabia) as a Post-Doctoral Researcher. In 2015, he joined the Electricity and Electronics Department, UPV/EHU as a Post-Doctoral Researcher and since 2017 he is an Adjunct Professor in the Electronic Technology Department (UPV/EHU). His areas of interest are experimental characterization of magnetic materials, in-circuit measurements and EM simulation.

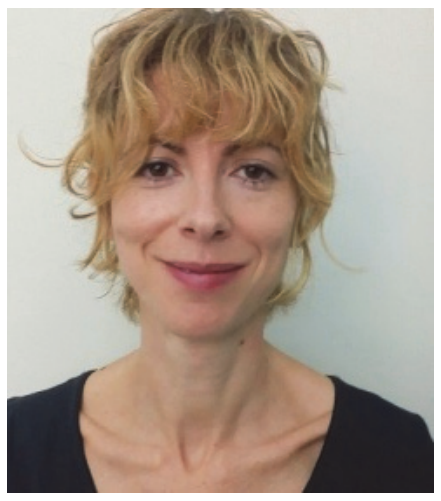

**Nerea Otegi** received the Ph.D. degree from the University of the Basque Country (UPV/EHU), Bilbao, Spain, in 2008. In 2002, she joined the Electricity and Electronics Department, UPV/EHU, where she has been an Associate Professor since 2006. Her areas of interest include noise characterization at microwave frequencies and nonlinear analysis of microwave circuits.

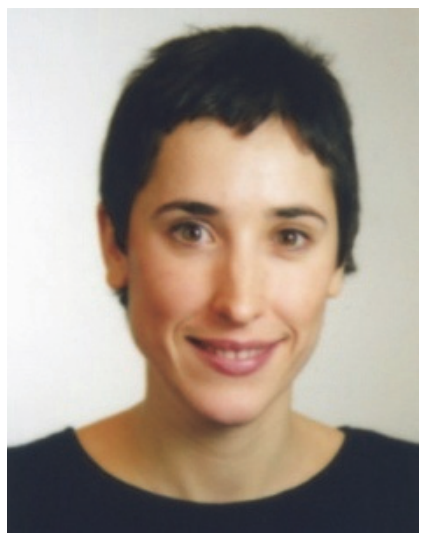

**Aitziber Anakabe** received the Ph.D. degree in electronics from the University of the Basque Country (UPV/EHU), Bilbao, Spain, in 2004. In 1999, she joined the Electricity and Electronics Department, UPV/EHU, where she was involved with the stability analysis of nonlinear microwave circuits. In 2004, she joined the French Space Agency (CNES), Toulouse, France, as a Post-Doctoral Researcher. In 2005, she rejoined the Electricity and Electronics Department, UPV/EHU, where, since 2005, she has been an Associate Professor. Her research deals with nonlinear analysis and modeling of microwave circuits and measurement techniques.

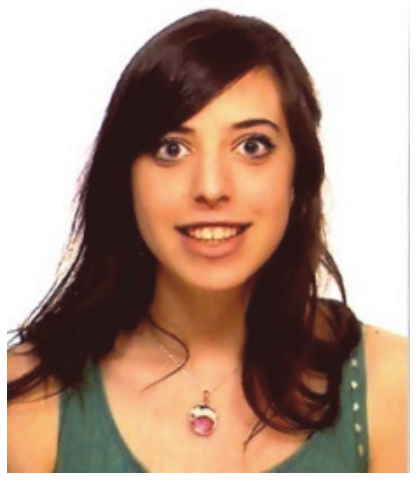

**Libe Mori** was born in Donostia, Spain, in 1992. She received de M.Sc. degree in electronics engineering from the University of the Basque Country (UPV/EHU), Bibao Spain, in 2015 and is currently working toward the Ph.D. degree in linear and nonlinear stability analysis of microwave power amplifiers at UPV/EHU. She is currently with the Electricity and Electronics Department, UPV/EHU. Her main research interests include analysis, modeling and design of microwave circuits.

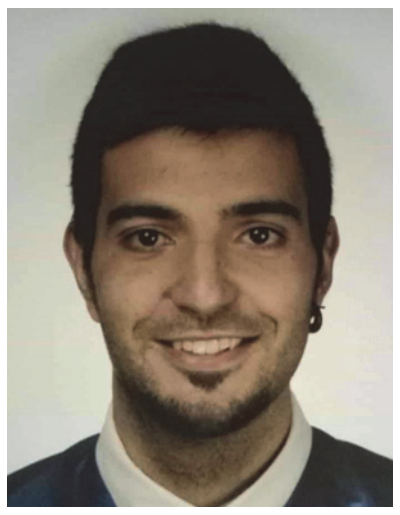

**Asier Barcenilla** was born in Bilbao, Spain, in 1994. He received de B.S in Electronics Engineering from the University of the Basque County (UPV/EHU), Leioa, Vizcaya, Spain, in 2017and he is currently working towards his M. Sc. in advanced electronic systems. He is currently with the Electricity and Electronics Department, UPV/EHU.

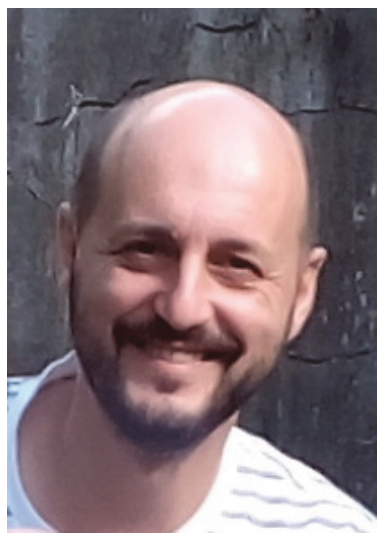

**Juan-Mari Collantes** received the Ph.D. degree in electronics from the University of Limoges, France, in 1996. Since February 1996, he has been an Associate Professor with the Electricity and Electronics Department, University of the Basque Country (UPV/EHU), Bilbao, Spain. In 1996 and 1998 he was an Invited Researcher with Agilent Technologies (formerly the Hewlett-Packard Company), Santa Rosa, CA. In 2003, he was with the French Space

Agency (CNES), Toulouse, France, where he was involved with power amplifier analysis, simulation, and modeling. His areas of interest include nonlinear analysis and design of microwave circuits, microwave measurement techniques and noise characterization.